\documentclass[review]{elsarticle}
\pdfoutput=1

\usepackage[hidelinks]{hyperref}
\usepackage[margin=2.5cm]{geometry}

\usepackage{pgfplots} 
\pgfplotsset{compat=newest} 
\pgfplotsset{plot coordinates/math parser=false} 
\newlength\figureheight 
\newlength\figurewidth 
\usepackage{amsmath,amsfonts, amssymb, mathtools, caption, subcaption}
\usepackage[font=small]{caption}
\usepackage{graphicx,wrapfig,lipsum}
\usepackage{tikzscale}
\usetikzlibrary{plotmarks}

\usepackage{multirow}

\usepackage{array}
\newcolumntype{K}[1]{>{\centering\arraybackslash}p{#1}}

\graphicspath{{Figures/}{../Figures/}}

\journal{European Journal of Operational Research}

\hypersetup{
  pdftitle={Gaussian Processes for Demand Unconstraining},
  pdfauthor={},
  plainpages=false,
  colorlinks=false,
}

\newcommand{\D}{\mathcal{D}}
\newcommand{\N}{\mathcal{N}}

\newcommand{\E}{\mathbb{E}}
\newcommand{\bff}{\mathbf{f}}
\newcommand{\bK}{\mathbf{K}}
\newcommand{\bmu}{\boldsymbol{\mu}}
\newcommand{\by}{\mathbf{y}}
\newcommand{\bx}{\mathbf{x}}
\newcommand{\be}{\mathbf{e}}

\newcommand{\deq}{\mathrel{\mathop:}=}

\begin{document}

\begin{frontmatter}

\title{Gaussian Processes for Demand Unconstraining}

\author[add1]{Ilan Price\corref{mycorrespondingauthor}}
\ead{ilan.price@maths.ox.ac.uk}
\cortext[mycorrespondingauthor]{Corresponding author}

\author[add1]{Jaroslav Fowkes\corref{}}
\ead{jaroslav.fowkes@maths.ox.ac.uk}

\author[add1,add2]{Daniel Hopman\corref{}}
\ead{daniel.hopman@maths.ox.ac.uk}

\address[add1]{Oxford-Emirates Data Science Lab, Mathematical Institute, University of Oxford, UK.}
\address[add2]{Vrije Universiteit, Amsterdam, Netherlands}

\begin{abstract}
	
	One of the key challenges in revenue management is unconstraining demand data. Existing state of the art single-class unconstraining methods make restrictive assumptions about the form of the underlying demand and can perform poorly when applied to data which breaks these assumptions. In this paper, we propose a novel unconstraining method that uses Gaussian process (GP) regression. We develop a novel GP model by constructing and implementing a new non-stationary covariance function for the GP which enables it to learn and extrapolate the underlying demand trend. We show that this method can cope with important features of realistic demand data, including nonlinear demand trends, variations in total demand, lengthy periods of constraining, non-exponential inter-arrival times, and discontinuities/changepoints in demand data. In all such circumstances, our results indicate that GPs outperform existing single-class unconstraining methods.
	
\end{abstract}


\end{frontmatter}



\section{Introduction}
\subsection{Demand Unconstraining for Revenue Management}

One of the key revenue management (RM) challenges which airlines, hotels, cruise ships (and other industries) all share is the need to make business decisions in the face of constrained (or censored) demand data \cite{talluri2006theory}. While we focus on the airline industry in this paper, our proposed methodology is directly applicable to the constrained demand problem in other industries.

Airlines commonly set booking limits on the number of cheaper fare-classes that can be purchased, or make cheaper fare-classes unavailable for booking at certain times in an attempt to divert some of that demand to the more expensive tickets still available. While a fare-class on a given flight route is available for booking, the demand for that `product', at that price, is accurately captured by its total recorded bookings. However, once the product has been unavailable for booking for a period of time, recorded bookings no longer capture true demand, and the demand data is said to be `constrained' or `censored'~\cite{yeoman2004revenue}. 

Practices which constrain demand data pose a big challenge for successful revenue management.  This is because many important decisions, including setting ticket prices, making changes to an airline's flight network, adding or removing capacity on a certain route, and many others, are all heavily dependent on accurate historical demand data. Moreover, precisely those decisions regarding which fare-classes to make unavailable (and for what periods of time) themselves depend on accurate demand data.

Since the mid-1990s, researchers have been studying ways to manage the constrained demand problem, and the proposed approaches fall under the banner of `unconstraining' methods. Broadly speaking, there are two types of product-model for which unconstraining methods are developed: single-class models which assume that demand for each fare-class is independent of the availability of (and demand for) all other fare-classes; and dependent demand models, where demand for a given fare-class on a given flight depends on the availability of (and demand for) other fare-classes on the same flight (or even other flights as well).

While assuming dependent demand might be theoretically appealing, multi-class methods have not been widely adopted in practice \cite{guo2012unconstraining}. The resulting consumer-choice based unconstraining methods can be much more complicated and expensive to incorporate at scale into airline revenue management systems, and rely on methodologies for estimating choice parameters and arrival rates which are as-of-yet ineffective \cite{guo2012unconstraining}. 

\begin{figure}
	\vspace{-0.5cm}
	\centering
	\includegraphics[width = 0.6\linewidth]{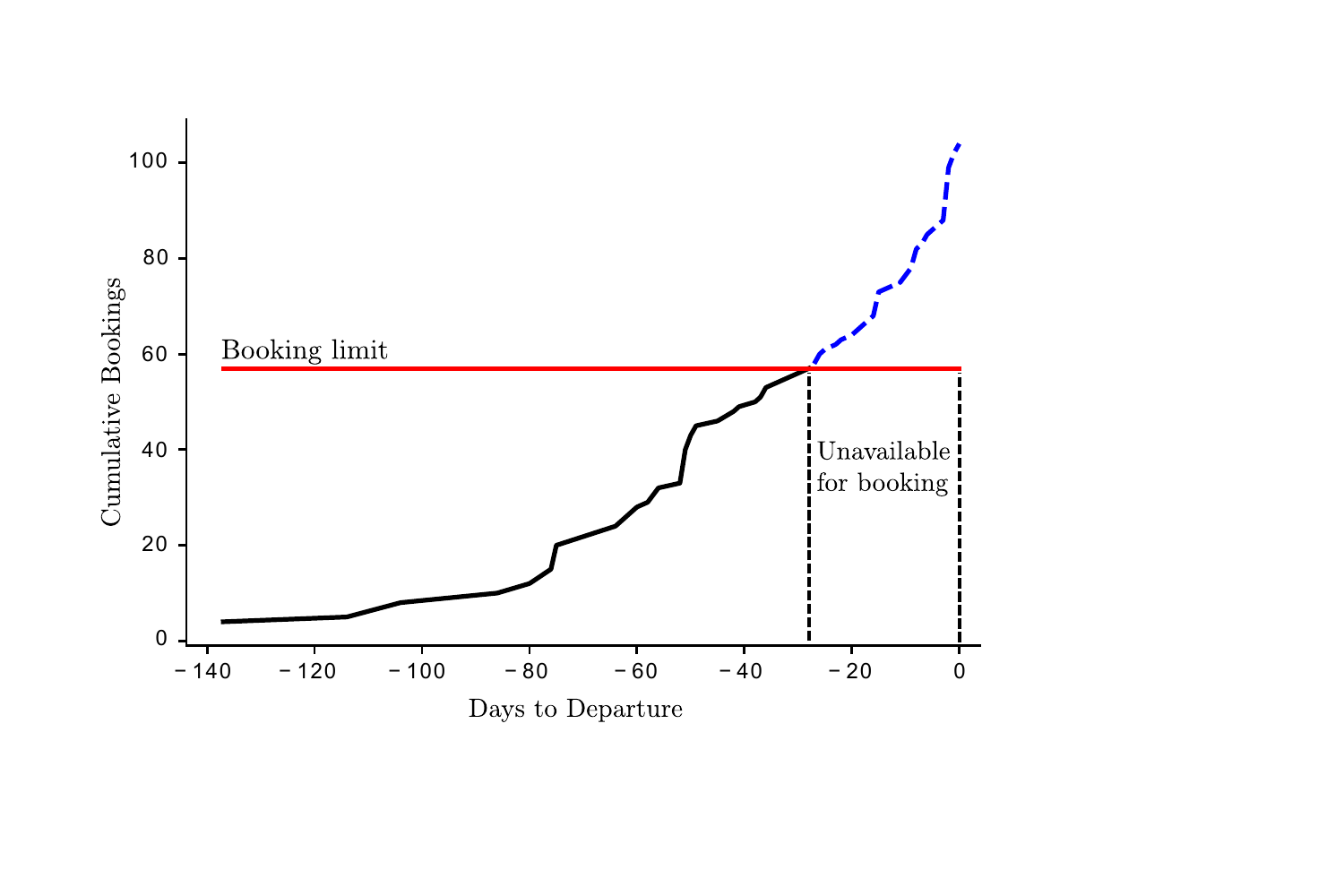}
	\caption{Schematic illustrating the single-class unconstraining problem. The demand curve shown represents cumulative bookings for a particular fare-class on a given flight. Observed bookings accurately capture true demand until the fare class becomes unavailable for booking due to an imposed booking limit, after which time the observed demand remains constant at the booking limit, even though the true demand exceeds that limit.}
	\label{Fig: Unconstraining Schematic}
\end{figure}

For these reasons, we focus in this paper on the single-class unconstraining problem, which can be stated as follows: given a demand curve which becomes constrained at some point, how do we accurately predict what demand would have been had no constraining occurred? We illustrate this problem in Figure~\ref{Fig: Unconstraining Schematic}, which highlights the difference between constrained and true demand as a consequence of an imposed booking limit.

\subsection{Mathematical Formulation of Demand Unconstraining}

For clarity and consistency across the different unconstraining methods, we mathematically formulate the unconstraining problem in quite general terms. 
Let $\D$ be the full set of demand observations under consideration (which may or may not be ordered, depending on the unconstraining method in question). We refer to demand observations, made when demand was not constrained in any way, as observations of \textit{true demand}, and those demand observations made in the presence of a constraint as observations of \textit{constrained demand}. We reserve the use of the term \textit{unconstrained demand} to refer exclusively to the \textit{output produced by an unconstraining method}, that is,  approximations of what the constrained demand observations would have been had there been no constraining.

We define $\D_T \subset \D$ as the subset of true demand observations, and $\D_C  \subset \D$ as the subset of constrained demand observations, such that $\D = \D_T \cup \D_C$ and $\D_T  \cap  \D_C = \emptyset$. We define $\D_U$ as the set of unconstrained demand values, corresponding to the unconstrained approximations of the elements of $\D_C$, and define $\widehat{\D} = \D_T \cup \D_U$, the full set of demand values where the constrained observations have been replaced by their unconstrained approximations. The demand unconstraining problem can therefore be stated mathematically as: use the available demand observations $\D = \D_T \cup \D_C$ to estimate the unconstrained demand values $\D_U$ as accurately as possible.  

\subsection{Gaussian Processes}
Gaussian processes (GPs) provide a robust statistical basis for inferring underlying statistical models from  observed data. GPs were first applied to time series analysis in 1949~\cite{wiener1949extrapolation}, but it was not until the 1970s that a general theory of GP prediction was developed \cite{o1978curve}. Since then, GPs have become a popular and very general framework for statistical modelling, and have been used to tackle a vast array of problems, including applications in machine learning~\cite{rasmussen2006gaussian}, atmospheric modelling \cite{fuentes2005model}, biochemical reactions \cite{gao2008gaussian}, and many others. 

One of the applications for which the use of GPs has yet to be investigated, however, is unconstraining demand. In this paper we propose and test a new GP model for use on this problem. We show that it outperforms state of the art unconstraining methods, coping much better with nonlinear and even discontinuous demand trends, variations in total demand, lengthy periods of constraining, and both exponential and non-exponential inter-arrival times.

\subsection{Paper Structure}

The rest of this paper is structured as follows. We review literature on single-class unconstraining methods in \autoref{Section: lit}. Next, we shift our focus to Gaussian processes: first, in \autoref{Section: GP regression}, we briefly introduce the mathematics of GP regression; we then move on, in \autoref{Section: GPs for unconstraining}, to motivate and develop the details of our GP model for application to the single-class unconstraining problem. 

In \autoref{Chapter: Experiments}, we describe three numerical experiments conducted to evaluate the performance of our proposed GP unconstraining method in comparison to state of the art alternatives. As a starting point, in \autoref{Section: Exp 1}, we repeat the most recent experiment from the literature comparing these methods. Second, we motivate for and design a modified version of this experiment with less restrictive assumptions used when constructing the test data, which we discuss in \autoref{Section: Exp 2}. Lastly, in \autoref{Sec: Exp 3}, we perform a third  experiment which uses generated data that more accurately resembles real Emirates Airlines bookings data.

Finally, in \autoref{Chapter: Changepoints} we extend our proposed unconstraining method to handle demand trends which exhibit kinks and discontinuities, illustrating its performance on three scenario-inspired test cases.


\section{Literature Review}\label{Section: lit}

Despite the large potential impact on profits, reviews of the RM literature show that the unconstraining problem has not received as much attention as perhaps it deserves \cite{bobb2008open, mcgill1999revenue, talluri2006theory}. Indeed, research on unconstraining only meaningfully began in the mid-1990s. In what follows, we categorise and give a brief overview of the main single-class unconstraining methods which have been proposed and compared since then. For more general reviews of existing unconstraining research, see~\cite{guo2012unconstraining, weatherford2016history}. 

The most rudimentary approaches to dealing with constrained data involve no mathematics at all. One such approach is to simply ignore the fact that the data is constrained, and another is to disregard the constrained data entirely, basing forecasts exclusively on true historical demand data.  The former approach, sometimes referred to as Na\"ive 1 or N1 \cite{guo2012unconstraining}, will of course lead to (possibly very large) underestimation of current and future demand, which can potentially cause a `spiral-down' in total revenue \cite{cooper2006models, guo2012unconstraining}.
The latter approach, sometimes referred to as Na\"ive 2 or N2 \cite{guo2012unconstraining}, may perform well in particular circumstances, for example, when only a very small number of data points are constrained, but in practice, this method can produce both significant over- and under-estimations of demand, depending on the context \cite{zeni2001improved}. 

All other methods in the literature that deal with constrained demand data employ a mathematical or statistical model for the purposes of unconstraining. 
We can divide the remaining methods into two categories based on an important conceptual difference in their approaches to unconstraining. Methods in the first category (which we term `\textbf{unordered methods}') are applied to a set of historical demand data from a group of past flights. These data points are treated as unordered, and the goal of the methods is to produce unconstrained estimates for the constrained elements of that data set. The second category (which we term `\textbf{time-series methods}'), consists of methods which are applied to one constrained demand curve at a time.  Time-series methods use demand data from a given flight up until the time that flight was constrained, to extrapolate what the actual demand for that flight would have been, had it not been constrained. 

The vast majority of existing unconstraining methods are unordered methods. One of the most elementary approaches in this category is known as mean-imputation, (or alternatively as Na\"ive 3, N3, or the `mixed approach'). This method involves simply comparing each constrained value with the mean of all unconstrained values, and replacing it with the larger of the two \cite{zeni2001improved}. Variations of this method use the median or some other specified percentile instead of the mean.

Salch proposed using the general statistical method of Expectation Maximisation (EM) for single class unconstraining in airline industry problems \cite{salch1997unconstraining}, and since then it has established itself as perhaps the most widely used unconstraining method. A variant of EM known as Projection Detruncation (PD) was first proposed for the purposes of unconstraining demand by Hopperstad \cite{hopperstad1995alternative}. Out of EM and PD, only EM has a rigorous statistical basis and has been proven to converge (under suitable assumptions \cite{gupta2011theory}) as  PD is based on a heuristic.

Skwarek \cite{skwarek1996competitive} proposed a method known as Pickup Detruncation, which calculates the amount of bookings made for a given flight over the period it was constrained as the average total bookings made in the same period before departure for flights that were not constrained. Around the same time, Wickham introduced the `Booking Profile' (BP) method \cite{wickham1995evaluation} (alternatively known as the `multiplicative method'). BP works by using historical (true) bookings data from different flights to build a bookings profile over time, from the day tickets go on sale until departure.

Van Ryzin and McGill applied the statistical method of Life Tables (LT) to unconstraining demand \cite{van2000revenue}. Unfortunately, the method tends to produce biased estimates \cite{lawless2011statistical}, and only produces unconstrained approximations of the mean and standard deviation, rather than estimates for each instance in which demand was constrained. Liu et al. \cite{liu2002estimating} proposed a method for use in the hospitality industry which uses parametric regression. It differs most notably from other methods in that it attempts to account for other demand-influencing factors when calculating the distribution of demand, such as length of hotel stay and competitors' room rates. 

The only existing method which falls decidedly into the `time-series methods' category was proposed by Queenan et al.\ \cite{queenan2007comparison}. They propose using the established forecasting algorithm known as Double Exponential Smoothing (DES), or ``Holt's Method'', for unconstraining demand, and compare its performance to EM, PD, LT and a variant of mean imputation. They report that, while in some cases EM outperforms DES, DES generally performs better on the most common booking curves shapes, and when the vast majority of the data is constrained. We discuss this paper in depth in \autoref{Section: Exp 1}.

Prior to Queenan et al.\ \cite{queenan2007comparison}, a number of other papers had been published comparing the accuracy and  revenue impact of many of the methods described above. Guo at al.\ \cite{guo2012unconstraining} report that \cite{skwarek1996revenue} and \cite{hopperstad1996passenger} compare the revenue impact of N2, N3, BP, and PD, finding that BP and PD outperform the Na\"ive methods. Weatherford~\cite{weatherford2000unconstraining} finds that out of EM, BP, N1, N2, and N3, the EM method best minimises the mean absolute error and best approximates the true mean of the data. Weatherford and Polt~\cite{weatherford2002better} and Zeni~\cite{zeni2001improved} compare EM, BP, N1, N2, N3, and PD, concluding that EM and PD are the best performing methods.
The primary take-home message from these comparisons is that EM, PD, and DES are the most competitive and widely-used single-class unconstraining methods developed so far.

\section{Gaussian Process Regression}\label{Section: GP regression}

The general idea behind Gaussian Process regression is very intuitive. We start by assuming a prior Gaussian distribution over functions, and then restrict our distribution to include only those functions which make sense given the observed data. More formally, our goal is to infer some unobserved (latent) function $f$ evaluated at a set of test inputs $X^* = \{ x_1^*,\dots, x_m^*\}$, using observed data $\by = \{y_1, \dots, y_n \}$ at points $X = \{x_1, \dots , x_n \}$. Let $\bff$ and $\bff^*$ be vectors of unobserved function values at inputs $X$ and $X^*$ respectively, and let $\theta_c$ be a set of covariance hyperparameters for $f$. 

The function $f(x)$ is a Gaussian process (GP) if, for every point $x$, $f(x)$ is a random variable, and for any finite set of points $\{x_1, x_2, \dots , x_n\}$, the set $\{ f(x_1), f(x_2), \dots , f(x_n)\}$ has joint Gaussian distribution whose mean and covariance are defined by a mean function $m(x)$ and covariance function $k(x,x')$ evaluated at the points $\{x_1, x_2, \dots , x_n\}$. We will assume throughout that the mean function is zero, as is standard in the literature. A covariance function takes the form of a kernel (or similarity) function mapping $x,x' \in X$ to~$\mathbb{R}$, which specifies the covariance between the random variables $f(x)$ and $f(x')$, denoted as
\begin{align*}
\text{Cov}[f(x'), f(x)] = k(x,x').
\end{align*}
Covariance functions are symmetric by definition and require that the covariance matrix $\bK_{i,j} \deq k(x_i,x_j)$ of the points $\{x_1, x_2, \dots , x_n\}$ must be Positive Semi-Definite (PSD) \cite{rasmussen2006gaussian}. A covariance function is said to be `stationary' if it is a function of $(x - x')$, making it invariant under translations in input space \cite{genton2001classes}. In contrast, when $k$ is not a function of $(x - x')$ the covariance function is known as `non-stationary'. 

Assuming a GP prior implies a joint Gaussian distribution over $\bff$ and $\bff^*$, and it is known \cite{rasmussen2006gaussian} that the conditional distribution of $\bff^*$ given $\bff$ is
\begin{align}\label{eq:posterior no noise}
\bff^* | \bff, X, X^*, \theta_c \sim \N (\bK_*^\top \bK^{-1}\bff, \;\bK_{**}- \bK_*^\top \bK^{-1}\bK_*), 
\end{align}
where the covariance matrices $\bK_{i,j} \deq k(x_i,x_j)$, $(\bK_*)_{i,j} \deq k(x^*_i,x_j)$, and $(\bK_{**})_{i,j} \deq k(x_i^*,x^*_j)$.

In general, the function $f$ is considered to be a latent function, meaning that we do not observe the actual function values $\bff$; rather we observe values $\by$ which are related to the true function values in a particular way. This relationship is defined by the observation model, that is, the likelihood of the observed values $p(\by|\bff,X,\theta_L)$ where $\theta_L$ denotes the set of likelihood hyperparameters. The specific form of the likelihood depends on the process one is trying to model and will not be Gaussian in general. 

Given our likelihood $p(\by | \bff, X, \theta_L)$ and our prior $p(\bff| X, \theta_c)$, we need to calculate the conditional posterior distribution $p(\bff | \by, X, \theta)$, where $\theta=\theta_c\cup\theta_L$. Bayes' rule, the cornerstone of Bayesian inference, allows us to compute this as follows:
\begin{align}\label{eq: conditional posterior} 
p(\bff | \by, X, \theta) = \dfrac{ p(\by | \bff, X, \theta_L)p(\bff| X, \theta_c) }{ \int p(\by | \bff, X, \theta_L )p(\bff| X, \theta_c) \text{d}\bff }.
\end{align}

In cases where the likelihood (observation model) $p(\by|\bff, X, \theta_L)$ is Gaussian, the marginal likelihood (the integral in the denominator) can be calculated exactly. In all other cases, the conditional posterior must be approximated. One standard approach is to construct a Gaussian approximation of the conditional posterior using the Laplace approximation \cite{rasmussen2006gaussian}, yielding
\begin{align}\label{eq:laplace1}
p(\bff | \by, X, \theta) \approx \N(\hat{\bff}, \Sigma^{-1}),
\end{align}
where the mode $\hat{\bff} \deq \arg\max_{\mathbf{f}} p(\bff | \by, X, \theta)$ and the precision matrix $\Sigma$ is the Hessian of the negative log conditional posterior evaluated at the mode:
\begin{align}\label{eq:laplace2}
\Sigma = -\nabla \nabla \log p(\bff|\by,X,\theta)\vert_{\bff = \hat{\bff}} = \bK^{-1} + \mathbf{W},
\end{align}
where $\mathbf{W}$ is the diagonal matrix with entries $W_{ii}= \nabla_{f_i}\nabla_{f_i}\log p(y_i | f_i,  x_i, \theta_L)\vert_{f_i = \hat{f}_i}$. For details of calculating the mode $\hat{\bff}$ and the precision matrix $\Sigma$ see \cite{rasmussen2006gaussian}.

Finally, to obtain the posterior predictive distribution, we combine the conditional probability \eqref{eq:posterior no noise} with the conditional posterior \eqref{eq: conditional posterior} and marginalise out the latent function values $\bff$: 
\begin{align*}
p(\bff^*|\by, X, X^*, \theta) = \int p(\bff^* | \bff, X, X^*, \theta_c) p(\bff | \by, X, X^*, \theta) \text{d}\bff.
\end{align*}
In cases where the conditional posterior has been approximated with the Laplace approximation (\ref{eq:laplace1}-\ref{eq:laplace2}), computing this integral gives
\begin{align}\label{eq: posterior pred}
\bff^*|\by, X, X^*, \theta \sim \N(\bmu_p,\bK_p),
\end{align}
where the posterior mean and covariance are given by
\begin{align}\label{eq: posterior mean and cov}
\begin{split}
\bmu_p &= \bK_*^\top \nabla \log p(\by| \bff, X, \theta_L)\vert_{\bff=\hat{\bff}},\\
\bK_p & = \bK_{**} - \bK_*^\top [\bK + \mathbf{W}]^{-1}\bK_*.
\end{split}
\end{align} 

Before we can make predictions using the posterior predictive distribution \eqref{eq: posterior pred}, we remove its dependence on $\theta$ by marginalising out the hyperparameters. To do this we need to compute 
\begin{align}\label{eq: post pred before bayes}
p(\bff^*|\by, X, X^*) = \int_\theta p(\bff^*|\by, X, X^*, \theta)p(\theta|\by ) \text{d}\theta.
\end{align}
We apply Bayes' rule to $p(\theta | \by)$ in order to transform the integral in \eqref{eq: post pred before bayes} into one in terms of the marginal likelihood $p(\by|\theta)$. The resulting equation is
\begin{align}\label{eq:post-pred integrals}
p(\bff^*|\by, X, X^*) = \frac{1}{Z}\int_\theta p(\bff^*|\by, X, X^*, \theta)p(\by|\theta )p(\theta) \text{d}\theta,
\end{align}
where $Z = \int_{\tilde{\theta}} p(\by|\tilde{\theta})p(\tilde{\theta}) \text{d}\tilde{\theta}$ is the marginalisation constant, and $p(\theta)$  
is a prior distribution over our hyperparameters which must be specified. Garnett et al. \cite{garnett2009sequential}  approximate this integral with Bayesian Monte Carlo techniques, while Saat\c{c}i et al.\ \cite{saatcci2010gaussian} recommend a simple quadrature approximation instead. We choose to implement the latter, approximating the integrals in \eqref{eq:post-pred integrals} with sums such that
\begin{align*}
p(\bff^*|\by, X, X^*) \approx \sum_{\theta_g} p(\bff^*|\by,X,X^*,\theta_g)\left( \frac{p(\by|\theta_g)}{\sum_{\theta_g}p(\by|\theta_g) }\right),
\end{align*}
where $\{\theta_g\}$ is a grid placed over a reasonable subspace of the GP hyperparameters, and where we have assumed a uniform prior probability mass at each grid point.

\section{Gaussian Processes for Unconstraining Demand}\label{Section: GPs for unconstraining}

\subsection{Problem Setup}
The single-class unconstraining problem, from a time-series perspective, is equivalent to a short-term forecasting or extrapolation problem. We propose using GP regression to learn the underlying booking trend for a particular flight from the bookings data up until the time it was made unavailable (i.e.\ from true demand observations), and to make predictions about what the true demand would have been thereafter. 

Unlike DES, which takes cumulative bookings as inputs and forecasts directly in `cumulative-space', our aim is to model the underlying booking trend, and hence we perform GP regression on daily bookings. We take the set of points $X$ to be the days from when tickets were made available until the day they were constrained, and the observed data $\by = \D_T$ to be the set of observed daily bookings on these days. Since we are in `daily-space', the constrained demand observations in $\D_C$ are all zeros, corresponding to inputs $X^*$ which are the remaining days before departure when booking was not possible. Once we have defined a suitable model, described in the remainder of this section, we follow the steps described in \autoref{Section: GP regression} to calculate the posterior predictive distribution $p(\bff^* | \by, X, X^*)$ and forecast our daily unconstrained demand values to be the mean of this posterior predictive distribution. This approach is illustrated in \autoref{Fig: gp method illustration} below.

\begin{figure} [h] 
	\centering
	\begin{subfigure}{0.5\textwidth}
		\centering
		\includegraphics[width=0.9\textwidth]{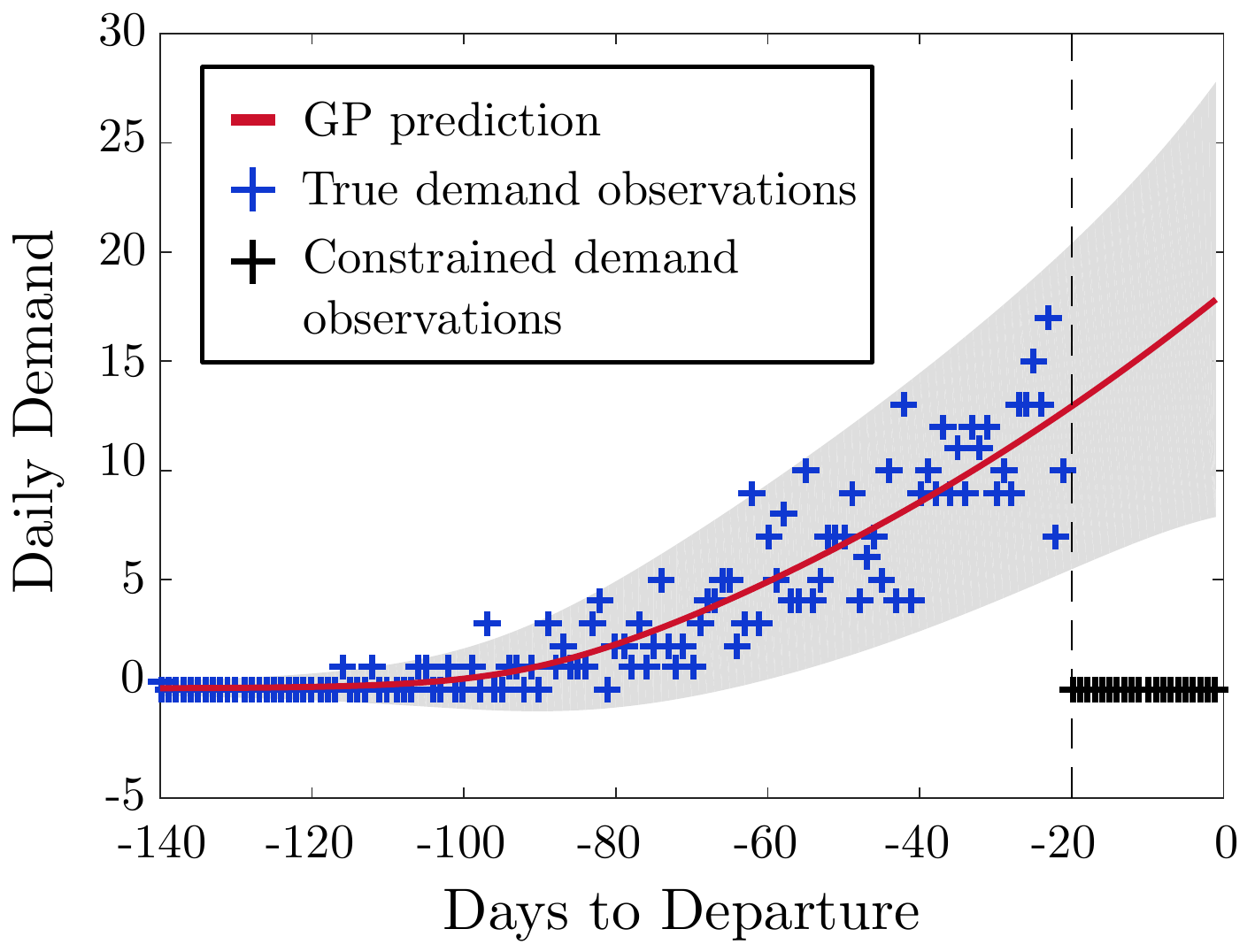}		
		\label{Fig: gp method illustration rs}
	\end{subfigure}%
	\begin{subfigure}{0.5\textwidth}
		\centering
		\includegraphics[width=0.9\textwidth]{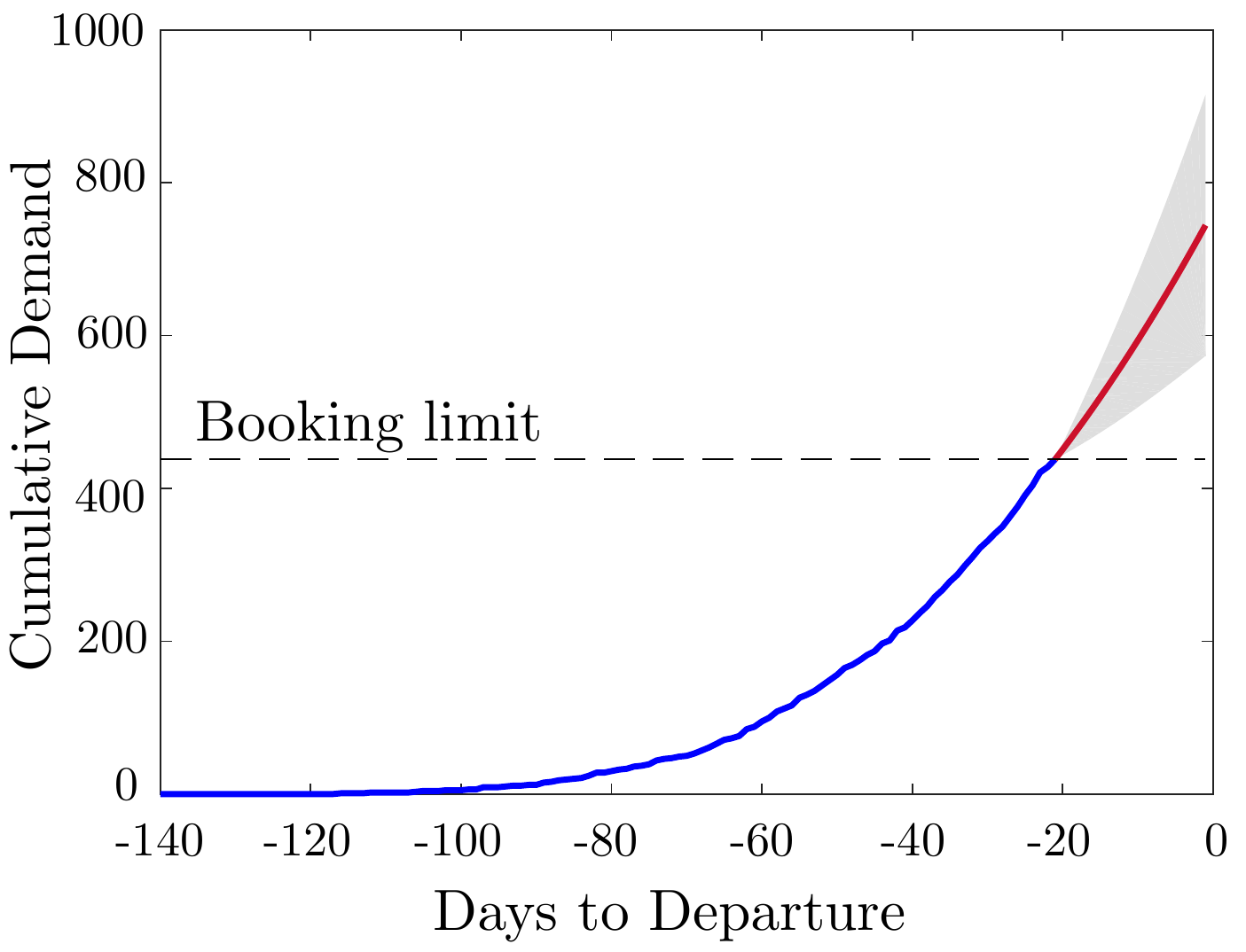}		
		\label{Fig: gp method illustration cumul}
	\end{subfigure}
	\caption{Illustration of GP regression for unconstraining demand. The figure on the left shows the mean prediction and confidence interval produced by our GP method, based on the true demand observations. The dotted black line indicates when the booking limit was reached, and the red line beyond this point shows the GP's unconstrained approximations. The figure on the right shows in red the reconstruction of the cumulative demand curve over the constrained period using the daily demand values predicted with the GP.}
	\label{Fig: gp method illustration}			
\end{figure}

\subsection{Motivation}	
Unordered methods, though currently favoured by airlines, face a number of key challenges.  Firstly, they require a significant amount of true historical demand data to perform well. Crucially, this must be data from flights which are assumed to have similar booking patterns and demand totals, since unordered methods all make the assumption that the demand data they use share a common underlying distribution. Since booking behaviours for flights vary with the month of departure, the weekday of departure, and even the time of departure, there needs to be a long history of accurate demand data for these methods to be applied. Not only is this a problem for new flight routes which do not yet have sufficient demand history; it is also a potential problem for popular fare-classes on peak-season flights which are almost always constrained at some point before departure. For these flights, while there is a long history of recorded demand data, very little of this will be true demand (as is necessary). This problem is clearly avoided by time-series methods, for which the only data needed to unconstrain a particular demand curve is the demand data from that curve prior to it being constrained.

A second problem for unordered methods is their inability to account properly for exogenous circumstances which change over time. Flight demand is affected by the strength of the economy, inflation and ticket prices, the relative strength of the origin and destination currencies, and many other factors. These are likely to vary over time, creating unaccounted-for variation in demand even among flights departing at the same time on the same weekday but in different months or years. Since unordered methods prioritise producing unconstrained demand estimates which are similar to past flights from different months and years, they cannot adequately take account of these exogenous effects. Time-series methods, on the other hand, implicitly consider these by utilising only the trend in demand for that specific flight up until it was constrained. 

For these reasons, we favour a time-series approach. DES, the only other distinctly time-series method, has a number of key limitations, the most important one being that it can only produce linear extrapolation. GPs, in contrast, have the ability to learn and extrapolate non-linear trends, which is an important advantage.

\subsection{Proposed Model}\label{Section: proposed model}

Our proposed GP model is based on the assumption that the flight bookings process (which can be thought of as an `arrivals process') is best modelled as an inhomogeneous Poisson point process, a standard assumption in RM \cite{talluri2006theory}. To build this assumption into our GP model, we use a Poisson likelihood for our observation model $p(\by|\bff,X,\theta_L)$, which is to say we assume the observed bookings on day $x_i$ to be a sample from a Poisson distribution. However, we cannot take $y_i \sim \text{Poiss}(f(x_i))$ since when $f(x_i)=0$, this is not a distribution. A standard approach \cite{rasmussen2010gaussian} is therefore to treat the number of bookings on day $x_i$  as a sample from a Poisson distribution with rate $\lambda(x_i) = \log (1+ \text{e}^{f(x_i)})$.

As mentioned in \autoref{Section: GP regression}, we use a zero mean function for the GP prior, and our choice of covariance function is influenced by two key considerations. The first is that we are going to be using our posterior distribution for the purposes of extrapolation, since we are predicting what the demand trend would have been beyond the time at which it was constrained. The second is that we assume the rate $\lambda(t)$ of the underlying inhomogeneous Poisson process is generally smooth (though we do not exclude the possibility of sudden, infrequent changes in the scale of (and/or trend in) demand, with which we deal explicitly in \autoref{Chapter: Changepoints}).

The extrapolation consideration is important since when performing GP regression using most stationary covariance functions, the posterior mean tends towards the prior mean as one moves further away from the observed data $\by$, making these covariance functions poor candidates for applications involving extrapolation. For better performance on extrapolation problems, Wilson et al.\ \cite{wilson2013gaussian} propose a spectral mixture covariance function, which uses a weighted product of multiple Gaussians in constructing the spectral density of a new stationary covariance function. However, while their results are impressive, their covariance function does not entirely avoid the `mean problem' faced by other stationary covariance functions, and furthermore its performance is highly sensitive to its hyperparameter initialisation, requiring a computationally expensive initialisation procedure in order to choose appropriate initial values.

\subsection{A Non-Stationary Covariance Function}
We propose a non-stationary covariance function for our GP model that does not suffer from the `mean problem' faced by most stationary covariance functions. Further motivation for a non-stationary covariance function comes from considering the bookings process we are attempting to model. 
Consider an inhomogeneous Poisson process with rate $\lambda(t)$. Let $B(t)$ be a random variable representing the total number of bookings in the window from time $0$ to time $t$, such that Var$[B(t)] = \E[B(t)] = \int_0^t \lambda(t) \text{d}t$ is the variance of $B(t)$.
For $s$ such that $0<s<t$, $B(s)$ and $B(t)-B(s)$ are independent and thus have a covariance of zero. This means we can write $\text{Cov}[B(s), B(t)] = \text{Cov}[B(s), B(s)] + \text{Cov}[B(s), B(t) - B(s)] = \text{Var}[B(s)]$,
which clearly shows that the covariance in `cumulative-space' is non-stationary.

Our proposed covariance function is motivated by the polynomial covariance function  \cite{rasmussen2006gaussian}:
\begin{align}\label{eq: fixed cov func}
k(x,x') = \sigma^2(x^\top x' + c)^p,
\end{align}
where $\theta_c = \{\sigma , c\}$ are hyperparameters and the degree $p$ is some specified positive integer.  For a given degree, GP regression with this covariance function can be shown to be equivalent to Bayesian polynomial regression \cite{rasmussen2006gaussian}. This serves as a sensible starting point for our model, since the `smoothness' assumption on the underlying Poisson intensity means that it can likely be well approximated with a polynomial. 

\subsection{Automatic Degree Inference}
Since we do not know a priori what the degree $p$ of the polynomial covariance function \eqref{eq: fixed cov func} will be, we propose a new covariance function which treats $p$ as a hyperparameter as well, to be inferred from the data like $\sigma$ and $c$. Our proposed covariance function is therefore also of the form
\begin{align}\label{eq: our cov func}
k(x,x') = \sigma^2(x^\top x' + c)^p,
\end{align}
but where this time $\theta_c = \{\sigma , c, p\}$ are the covariance hyperparameters. 

Once $p$ is a hyperparameter, it cannot be restricted to integer values. Polynomial kernels with fractional degree are not unprecedented, however. Kernels of the form
$(x^Tx')^p$, for $0<p<1$
have been used before for facial recognition using the kernel PCA method \cite{liu2004gabor}, and Rossius at al.\ \cite{rossius1998short} discuss the use of kernel functions of the form
$(x^Tx' + 1)^p$,
in Support Vector Machines, and the impact of non-integer values of $p$ when $x^Tx' <-1$, in which case the base raised to a non-integer power is negative. In both cases, however, $p$ was considered to be a fixed (albeit non-integer) value and we propose generalising the GP regression framework by letting $p$ be a covariance function hyperparameter which is automatically inferred from the data. To the best of our knowledge this has never been done before.

With our proposed covariance function \eqref{eq: our cov func} the covariance matrix $\bK$ becomes
\begin{align*}
\bK = \sigma^2(\bx \bx^\top + c \be \be^\top)^{\diamond p},
\end{align*}
where $\be$ is the vector of ones and $\cdot^{\diamond p}$ denotes a Hadamard power (the exponent applied element-wise). Recall from \autoref{Section: GP regression} that the covariance matrix $\bK$ is required to be Positive Semi-Definite (PSD). 
We prove that as long as we ensure that all $x, x'\geq 0$, and $c>0$, the rank 2 matrix $\bx \bx^\top + c \be \be^\top$ is PSD (we include the proof in the Supplementary Material for completeness).

It has further been shown that if a matrix $A\in \mathbb{R}^{n\times n}$ is PSD, then whether or not $A^{\diamond p}$ can be guaranteed to be PSD depends on the rank of $A$ and the value of $p$ \cite{guillot2015complete}. Fitzgerald at al.\ \cite{fitzgerald1977fractional} prove that if Rank$(A)\geq 2$, $A^{\diamond p}$ is only PSD if $p \in \mathbf{N} \cup [n-2, \infty)$.  Since in our case $n$ is the total number of training data points (the elements of $\D_T$), the inferred degree is very unlikely to be greater than or equal to $(n-2)$. We can therefore conclude that our proposed covariance function unfortunately does not, in general, result in a PSD matrix $\bK$. However, it is not uncommon for non-PSD kernels to be used nonetheless in applications where they perform well \cite{liu2004gabor}. We therefore adopt the common strategy \cite{rigdon1997not} of adding a sufficiently large perturbation to the spectrum of $\bK$, such that its indefiniteness is no longer a problem. Though this `artificial' shift causes bias in the resulting predictions, we do not find this to be an issue in practice. 

In fact, there is an intuitive way of interpreting the bias introduced by this shift. From \eqref{eq: posterior mean and cov} in \autoref{Section: GP regression}, and using the fact that $\nabla \log p(\by|\bff)\vert_{\bff=\hat{\bff}} = \bK^{-1}\hat{\bff}$ \cite{rasmussen2006gaussian} we see that without a shift, our posterior predictive mean would be given by
$\bmu_p = \bK_*^\top \bK^{-1}\hat{\bff}$.
When we shift the spectrum of $\bK$ by adding some diagonal matrix $\mathbf{D}$, this becomes 
\begin{align}\label{eq: post mean with shift}
\bmu_p &= \bK_*^\top (\bK+\mathbf{D})^{-1}\hat{\bff}.
\end{align}

Now let us compare this with the posterior predictive mean produced by an unshifted model, which instead uses a Gaussian likelihood, that is, assuming that observations include some noise which is normally distributed, $y_i \sim f(x_i) + \varepsilon$, where, $\varepsilon \sim \N(f_i, \sigma_n^2)$. In this case, the posterior predictive mean is 
\begin{align*}
\bmu_p &= \bK_*^\top (\bK+\sigma_n^2 I)^{-1}\by,
\end{align*}
which is very similar to the form of \eqref{eq: post mean with shift} with $\by$ having replaced $\hat{\bff}$, and $\sigma_n^2 I$ replacing $\mathbf{D}$. In other words, we can understand the added shift $\mathbf{D}$ as adding an implicit assumption of a certain noise level in the data. In our case, we scale our inputs so that $x_i \in [0,1]$, and use $\mathbf{D}=I$.

\subsection{Implementation}\label{subsec: implementation}
To implement the GP regression method described in this section, we build upon the existing GPML MATLAB library created by Rasmussen and Williams \cite{rasmussen2010gaussian}. The library is already well equipped with most GP functionality, and is modular, such that functions for the different components of GP regression are defined independently, making it possible to incorporate new features into the existing library.  We extend GPML in two ways: first, we develop a new covariance function file to implement the variable degree polynomial covariance function defined in \eqref{eq: our cov func}; second, we develop code to implement a quadrature method to approximate the integral given in \eqref{eq:post-pred integrals}, which is required to marginalise the hyperparameters. However, since this code is not currently vectorised, the computational time (especially for the quadrature approximation) is significantly longer than it could otherwise be (between 5 and 20 seconds depending on the fineness of the quadrature grid, with an AMD FX 4350 Quad-Core Processor).

\section{Numerical Experiments}\label{Chapter: Experiments}

\subsection{Experiment 1: `The Queenan Framework'}\label{Section: Exp 1}

\begin{figure} [h] 
	\centering
	\begin{subfigure}{0.5\textwidth}
		\centering
		\includegraphics[width=0.9\textwidth]{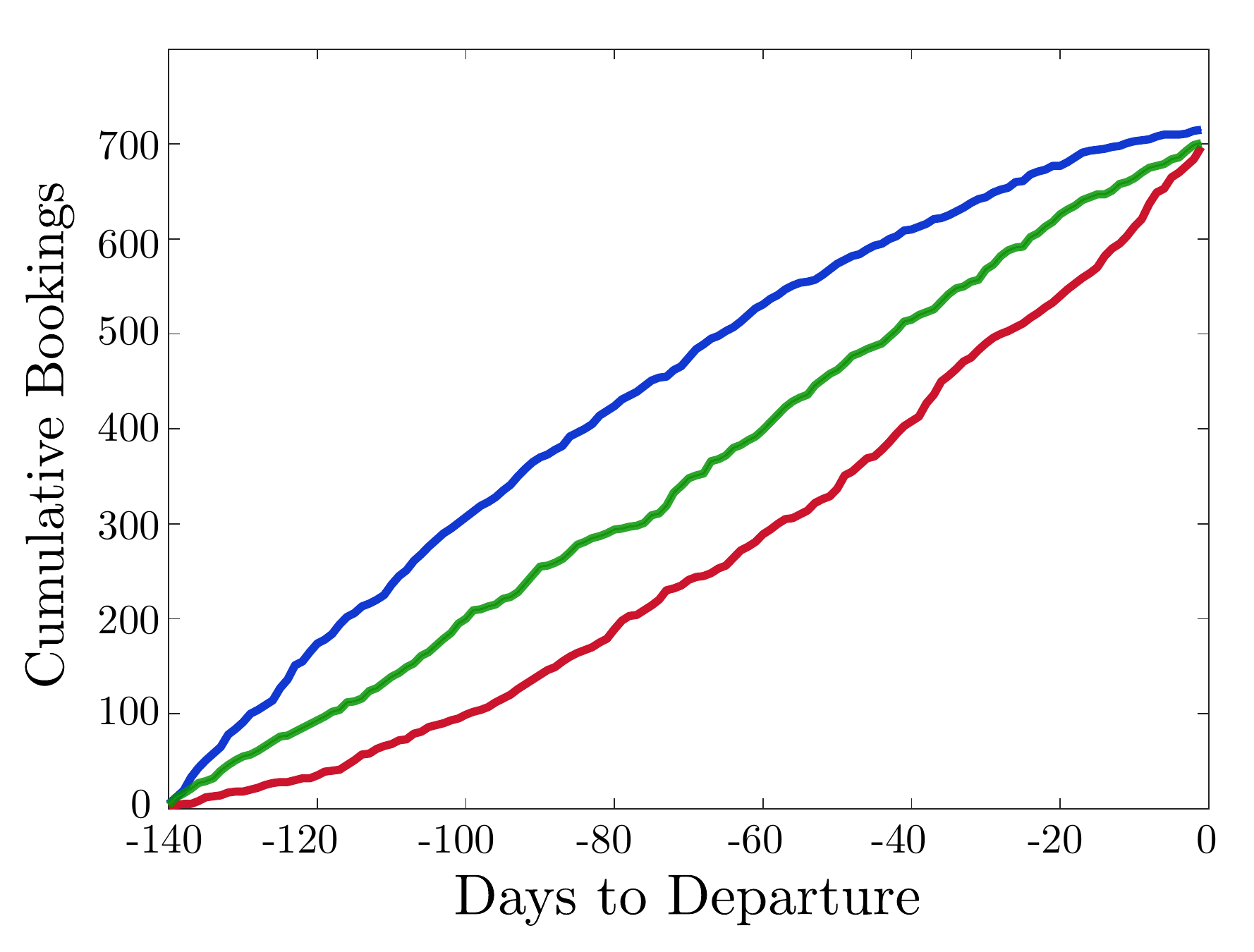}		
		\caption{Experiment 1}
		\label{Fig: Queenan curve types}
	\end{subfigure}%
	\begin{subfigure}{0.5\textwidth}
		\centering
		\includegraphics[width=0.9\textwidth]{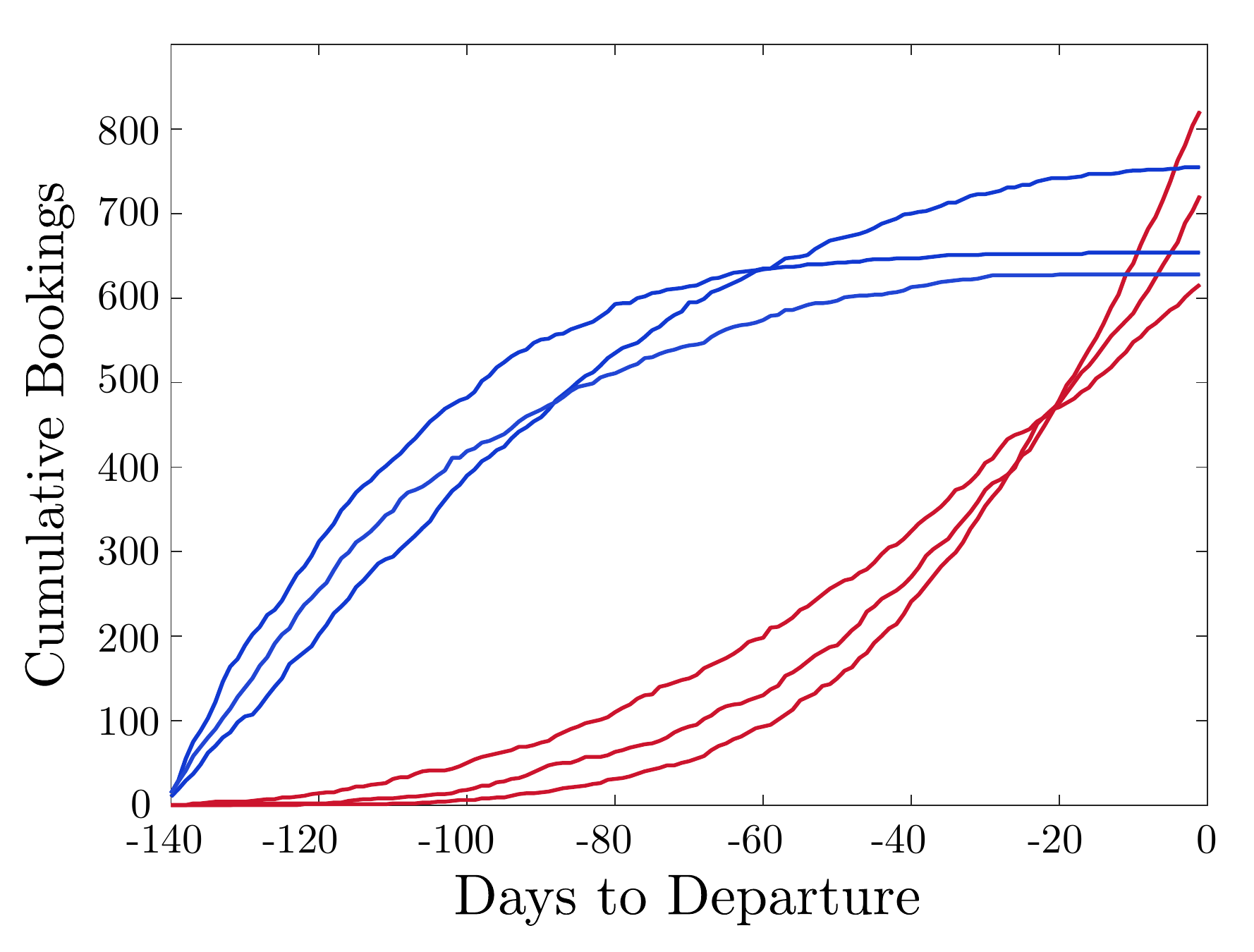}
		\caption{Experiment 2}
		\label{Fig: Generlised Queenan curves}
	\end{subfigure}
	\caption{Sample demand curves generated in Experiments 1 and 2. Figure (a) shows a sample from the convex, concave and homogeneous curve sets (red, blue, and green, respectively) which are generated in Experiment 1. All 100 curves in each set have very similar shapes to the curves shown here. Figure (b) shows three samples from the convex and concave curve sets (red and blue, respectively) which are generated for Experiment 2. This illustrates the more realistic variation in the extent of convexity/concavity among curves used in Experiment 2.}
	\label{Fig: Curve Examples}			
\end{figure}

We begin by reproducing the only experiment in the literature comparing the performance of the three best performing methods --- DES, EM, and PD --- presented by Queenan at al.\ \cite{queenan2007comparison}. They propose three typical cumulative demand curve types --- convex, concave, and homogeneous (approximately linear), examples of which are shown in \autoref{Fig: Queenan curve types} --- and compare the performance of DES, EM, PD, LT, and N3 for the three curve types. In this section, we reproduce their experiment for DES, EM, PD (as well as the variants `EM Daily' and `PD Daily' described below) and compare these to our proposed GP regression unconstraining method. 

\subsubsection{Constructing the Test Curves}

For each of the three curve types, 100 cumulative demand curves are generated, each with 140 data points, running from 140 days before departure up until the day before departure. 
To create one convex curve, daily demand for the first twenty days (140 -- 121 days before departure) is sampled from a Poisson distribution with $\lambda = 2$. For the next twenty days, daily demand is sampled from a Poisson distribution with $\lambda = 3$, and so on, such that the final twenty days before departure have daily demand sampled from a Poisson distribution with $\lambda = 8$. This process is repeated 100 times to create the 100 convex curves. The creation of the 100 concave curves follows a similar procedure, the only difference being that the mean of the Poisson distribution begins instead at $\lambda=8$ for the first 20 days, and decreases by 1 every 20 days, such that demand over the 20 day period before departure is sampled from a Poisson distribution with $\lambda = 2$. In the homogeneous/linear case, daily demand is sampled from a Poisson distribution with $\lambda = 5$ for all 140 days before departure. We note that this process for generating test curves is equivalent to simulating a piecewise-homogeneous Poisson process, where the inter-arrival times within each day are exponentially distributed. 

The next step, for each curve type, is to calculate a set of 100 random booking limits, corresponding to each of the 100 curves. Those curves whose cumulative demand exceeds their corresponding generated booking limit are the constrained curves. 
In each case, booking limits are generated five times, in such a way that an increasing proportion of curves are constrained each time: the first set of booking limits constrains approximately 20\% of the curves, the second set constrains approximately 40\% of curves, the third constrains 60\%, the fourth constrains 80\%, and finally the fifth set constrains 98\% of curves (see the Supplementary Material for details). These booking limits are used to artificially constrain the relevant booking curves. The various unconstraining methods are then applied to this artificially constrained data, producing unconstrained approximations, which are then compared with the `true' generated data to evaluate their performance. 

\subsubsection{Applying the Unconstraining Methods}
In order to apply the unconstraining methods, we first need to construct our set of demand data $\D =\D_T \cup \D_C$, using the curves and booking limits generated as described above.

For EM and PD this is done as follows: the set of actual cumulative demand totals for a given curve type is given by $\mathcal{A} = \{a_1, \dots, a_{100}\}$, where $a_i$ is the total cumulative demand of the $i^{th}$ curve. For each set of booking limits $\mathcal{B}$, there is some subset $\mathcal{A}_C \subset \mathcal{A}$ containing those demand totals which are greater than (and therefore constrained by) their corresponding booking limits. We set $d_i = b_i$ for all constrained observations, and $d_i = a_i$ otherwise, which gives us our set $\D$, to which EM and PD\footnote{We apply PD using $\tau=0.5$.} are applied (see the Supplementary Material for details). Note therefore, that standard EM and PD are only applied to the cumulative bookings totals the day before departure.

It is also possible, however, to  instead apply EM and PD to the daily bookings data from each day that any flight was constrained, thereby producing unconstrained approximations of the daily bookings for all constrained flights. Adding these approximations of bookings made on each constrained day to the bookings made prior to constraining, yields the total cumulative unconstrained approximation for that flight. We term this variation of the method `EM Daily' (and `PD Daily').  For EM and PD Daily, the procedure for assembling $\D$ is slightly different. The first step is to identify the first day that any curve in the set becomes constrained (that is, the earliest any curve exceeds its booking limit). We call this day $t_{\text{max}}$. Next, a separate set $\D=\D_T\cup\D_C$ is created for each day $t_k$, starting from $t_{\text{max}}$ up until departure. On each of these days, $\D_T$ contains the daily bookings from those curves whose cumulative totals by that day are still below their corresponding booking limit $b_i$. The set $\D_C$ contains all zeros (one for each curve which has surpassed its booking limit $b_i$ by day $t_k$).

In the case of the time-series methods, DES and our proposed GP regression method, each constrained curve is unconstrained independently. For each constrained curve, we calculate how many days before departure the booking limit was reached. In the case of DES, we then define $\D_T$ to contain the cumulative bookings up until that day, and all elements of $\D_C$ to be equal to the booking limit for that curve. Together these give us $\D$ and DES is applied to approximate $\D_U$ (see the Supplementary Material for details). For GP regression, we define $\D_T$ to contain the daily bookings up until the day of constraining, and all elements of $\D_C$ to be equal to zero. GP regression is applied in each case as described in \autoref{Section: GPs for unconstraining}. All four of DES, GP, EM Daily and PD Daily are therefore used to unconstrain data from the whole constrained period leading up to departure.

\subsubsection{Results}
Given the stochastic nature of the experiment, we repeat it five times and average the results. We present these results in \autoref{table: litmeanresults}, which mimics the format of those reported by Queenan et al.\ \cite{queenan2007comparison}. The success of each method is judged by calculating the mean of the set $\widehat{\D}= \D_T \cup \D_U$ containing true and unconstrained demand values, and comparing this with the mean of the actual demand set $\mathcal{A}$. We call this the E1 error, given by E1 $ = 100 * \left(\E[\widehat{\D}] - \E[\mathcal{A}]\right) / \E[\mathcal{A}] $.  

The results in \autoref{table: litmeanresults} show that when tested on the sets of convex and homogeneous curves, DES outperforms both EM and PD, though in the case of EM the margin is only significant when 98\% of curves are constrained. This is unsurprising as it is well known that EM performs poorly when almost all data is constrained. It is also apparent that as the proportion of constrained data increases, the performance of PD deteriorates faster than that of EM. Once again, it is to be expected that these results diverge, because with less data points the conditional mean  (used in EM) becomes less likely to be well approximated by the conditional median (used in PD). 

\begin{table}[h!]
	\centering
	\small
	\begin{tabular}{p{2.9cm}K{2.2cm}K{2.2cm}K{2.2cm}K{2.2cm}K{2.2cm}}
		\hline
		&\multicolumn{5}{c}{Proportion of Days Constrained}  \\ \hline
		& 20\%   & 40\%   & 60\%   & 80\%   & 98\%   \\ \hline
		\multicolumn{6}{c}{Convex}                          \\ \hline
		EM       & 0.05 & 0.17 & 0.40 & 0.36 & 1.44 \\
		PD       & 0.22 & 0.54 & 0.77 & 1.36 & 2.79 \\
		\textbf{DES}      & \textbf{0.03} & \textbf{0.07} & \textbf{0.06} & \textbf{0.06} & \textbf{0.21} \\
		EM Daily & 0.05 & \textbf{0.07} & 0.09 & 0.14 & 0.28 \\
		PD Daily & 0.06 &\textbf{0.07} & 0.09 & 0.14 & 0.31 \\
		GP      & 0.06 & 0.20 & 0.23 & 0.31 & 0.42 \\ \hline
		\multicolumn{6}{c}{Concave}                         \\ \hline
		EM       & 0.14 & 0.18 & 0.31 & 0.59 & 1.23 \\
		PD       & 0.29 & 0.48 & 0.74 & 1.55 & 3.52 \\
		DES      & 0.42 & 0.70 & 1.26 & 2.12 & 4.52 \\
		EM Daily & 0.06 & 0.12 & 0.15 & 0.25 & 0.48 \\
		\textbf{PD Daily} & \textbf{0.03} & 0.04 & \textbf{0.08} & \textbf{0.08} & \textbf{0.31} \\
		GP      & 0.06 & \textbf{0.02} & 0.12 & 0.30 & 0.52 \\ \hline
		\multicolumn{6}{c}{Homogeneous}                     \\ \hline
		EM       & 0.11 & 0.41 & 0.23 & 0.20 & 1.40 \\
		PD       & 0.20 & 0.47 & 0.87 & 1.46 & 2.86 \\
		DES      & \textbf{0.03} & 0.07 & 0.04 & \textbf{0.08} & 0.12 \\
		EM Daily & 0.04 & \textbf{0.05} & 0.08 & 0.17 & 0.29 \\
		PD Daily & 0.04 & 0.06 & 0.10 & 0.18 & 0.38 \\
		\textbf{GP}      & 0.04 & \textbf{0.05} & \textbf{0.03} & 0.09 & \textbf{0.07}   \\ \hline
	\end{tabular}
	\caption{ Percentage mean error (E1) for Experiment 1. The numbers indicate the percentage error in the mean of the set $\widehat{\D}$ produced by each method (with the best performing in bold). GP regression is either comparable to or better than EM, PD, and DES, but according to this measure performs slightly worse in general than Daily EM and PD.}
	\label{table: litmeanresults}
\end{table}

One noteworthy observation is that unlike all other methods, DES performed significantly worse on the concave set than on the convex set. The reason for this is that a given booking limit will constrain a concave curve for a longer period than a convex curve with the same total demand. Since DES only extrapolates linearly, it performs poorly on non-linear curves constrained for long periods of time.

The top three best performing methods across all curve shapes in this experiment are the daily variants of EM and PD, and our proposed GP regression method. All three produced very low percentage mean errors, which remained at or below 0.5\% even when 98\% of curves are constrained. In this experiment, GP is slightly outperformed by the other two. However, as we will see, this is more a result of unrealistic consistency across the curves in the test set than a reflection of the merits of daily EM and PD.

\subsubsection*{Assumptions behind the test curves}

There are three important assumptions which underpin the creation of the curve test set described above.
\begin{enumerate}
	\item The underlying `arrivals' or bookings process for all curves can be modelled with either (i) a constant underlying Poisson rate $\lambda$ or (ii) a linear $\lambda(t)$ approximated by a piecewise constant rate $\lambda$.
	\item All curves of a given type are described by precisely the same (piecewise) constant Poisson rate $\lambda$. 
	\item The variance in the total demand among all curves will be quite small (for example, the mean total demand for the convex curves was 696.74 with a standard deviation of only 27.93).
\end{enumerate}
It is in large part a consequence of creating booking curves based on these assumptions that EM and PD perform as well as they do. Indeed, the resulting data set to which EM and PD are applied is by construction almost exactly normally distributed. Recall that EM and PD are based on the assumption that the underlying distribution of the data is normal, and it should therefore come as no surprise that they perform well when tested on a dataset which has, by construction, a distribution that is so close to normal. 

However, the three assumptions listed above are all quite unrealistic. Firstly, our analysis of bookings data from Emirates Airlines shows that daily bookings~are~generally not well represented by a homogeneous Poisson process. Secondly, while modelling bookings as an inhomogeneous Poisson process is commonplace and sensible, the experiment allows for only a (crude approximation of a) linearly increasing or decreasing Poisson intensity, and fails to model non-linearly changing underlying Poisson rates $\lambda(t)$.

Thirdly, it is extremely unlikely that an airline would be able to isolate (a priori) a set of historical booking curves for which the bookings follow exactly the same underlying inhomogeneous Poisson process. It is much more likely that even for flights with the same general demand shape (e.g. convex booking curves), the change in Poisson intensity over time will vary --- it might be linear for some flights, and perhaps approximately quadratic or cubic for others, for example.  

Finally, even after restricting the set of flights under consideration to those that are believed to have similar booking trends historically, it is unlikely that the variance of total demand in this set would be so small (in the Emirates Airlines data set in \autoref{Sec: Exp 3}, the standard deviation was approximately 20\% of the mean total demand). The consequence of a test set with such a small variance is that EM continues to perform adequately even when curves are constrained for a long period of time, simply because it collapses into mean imputation (as shown in \autoref{Section: exp 2 results}) and the mean is mostly within two standard deviations of each instance of total demand.  

Given that these assumptions are unlikely to hold in realistic settings, the performance of EM and PD in this experiment should not be taken as evidence of their success in the airline industry.

\subsubsection*{Length of Constrained Period}

Queenan at al.\ follow Weatherford and Polt \cite{weatherford2002better} in their decision to set the artificial curve-specific booking limits based on what proportion of the demand curves they wish to constrain. However, the method they use to set booking limits to achieve this goal ends up constraining most curves for only a few days before departure --- even when 98\% of convex curves were constrained, the average length for which they were constrained was approximately 8 days.

Over periods as short as this, these cumulative demand curves can be fairly well approximated as linear, even though the overall trend might be decidedly non-linear. Since DES produces a linear extrapolation, it naturally performs well in this experiment on homogeneous and convex curves. However, in reality demand may well need to be unconstrained for a period over which the curve cannot be well approximated as linear. Such scenarios are tested by the concave case in this experiment, and unsurprisingly in these cases DES performs much more poorly than the other methods. To properly evaluate DES, the method for setting booking limits should therefore focus on the length of the constrained period rather than exclusively on what proportion of demand curves are constrained.

\subsubsection*{Metric for comparing methods}

To obtain the results presented in \autoref{table: litmeanresults}, we record the mean of the set $\widehat{\D}$ and compare it with the mean of $\mathcal{A}$, calculating the percentage error. However, given that one purpose of unconstraining is to predict what the true demand would have been for a particular flight whose demand was constrained, perhaps a more appropriate metric is to measure the difference between the unconstrained value $\widehat{d}_i$ and the actual value $a_i$, and consider the best method to be the one which minimises these absolute errors. 

Crucially, the method rankings produced by using these two metrics are not necessarily the same. For example, when judged according to which method most accurately reproduced the actual mean on convex curves with 98\% of them constrained (E1 error, \autoref{table: litmeanresults}), DES outperforms GP with an E1 error of 0.21\% as compared with 0.42\%. However, if instead we consider the average absolute error in the unconstrained approximations of final cumulative demand (which we call the E3 error), GP performs very similarly to DES (and even slightly outperforms it). The full set of E3 error results from Experiment 1 are shown in Table 1 in the Supplementary Material. Since accurately unconstraining each instance of constrained demand is a crucial function of unconstraining methods, considering only the percentage mean error compromises the assessment of the benefits of using each method in practice.

\subsection{Experiment 2: Generalised Queenan}\label{Section: Exp 2}
In this section we design and conduct a modified version of Experiment 1, creating sets of curves based on less restrictive assumptions, and use multiple metrics for adjudicating the relative performance of each method. We focus on the case of convex and concave curves, since truly homogeneous curves are much easier to unconstrain.

\subsubsection{Constructing the Test Curves}
In both the convex and concave case, we construct 90 demand curves, each with 140 data points (one for each day in the lead up to departure). The number of bookings $d_t$ (on day $t$ before departure) is sampled from a Poisson distribution with rate $\lambda(t)$. Instead of assuming the same underlying (piecewise constant) trend in $\lambda$ for every curve, however, we model 30 curves as having a linearly changing $\lambda(t)$, 30 curves with a quadratic $\lambda(t)$, and the last 30 curves with a cubic $\lambda(t)$. The resulting set of test curves has a similar mean demand to those in Experiment 1, but is more realistic in two key ways: 1) there is greater variance in total cumulative demand among the curves (a standard deviation of approximately 65 (instead of 28) for convex curves, and 67 (instead of 27) for concave curves), and 2) there is more variation in the shape of the demand curves (see \autoref{Fig: Generlised Queenan curves} above).

We randomly select 15 out of each set of 30 curves to constrain, such that a total of 45 out of 90 curves of each shape are constrained. We repeat this process three times, constraining the curves for 5, 10 and then 20 days prior to departure. We use this process instead of the booking limits procedure followed by Queenan at al.\ because it enables us to control the length of the constrained period, ensuring it is sufficiently long, without constraining 100\% of the curves (in which case we could not apply EM and PD). Our results show that it is not necessary to constrain as much as 80\% or 98\% of curves to illustrate the problems with EM and PD, and we restrict our attention to the case when only half the curves are constrained. As in \autoref{Section: Exp 1}, we repeat the whole experiment five times and average the results. 

\subsubsection{Results}\label{Section: exp 2 results}
\autoref{table: exp 2 results} summarises the results of this second experiment. We report three different error measures for the sake of completeness, each with their own motivation. As before, we include the percentage mean error (E1), the error type reported in \cite{queenan2007comparison} and \cite{weatherford2002better}, which is a measure of how close the mean of $\widehat{\D}$ is to the actual mean. Next, we include the average absolute daily error (that is, the average distance between the actual cumulative demand curve and the unconstrained curve during the constrained period), which we call E2. This indicates which method most accurately reproduces the constrained portion of the true demand curve, an important metric from the point of view of various airline applications (note that E2 cannot be measured for standard EM and PD because they are not applied to any data except the cumulative demand the day before departure). Finally, we report the average absolute difference between the unconstrained approximation of the final cumulative demand and the actual final cumulative demand of the constrained curves (E3).

\begin{table}[h!]
	\centering
	\small
	\begin{tabular}{p{3.3cm}K{1cm}K{1cm}K{1cm}|K{1cm}K{1cm}K{1cm}|K{1cm}K{1cm}K{1cm}}
		\hline
		& \multicolumn{9}{c}{No. of Days Constrained}                                              \\ \hline
		& \multicolumn{3}{c|}{5 Days} & \multicolumn{3}{c|}{10 Days} & \multicolumn{3}{c}{20 Days} \\ \hline
		Convex & E1 & E2  				&  E3          			& E1          	  & E2    			& E3       			& E1            & E2     & E3      \\ \hline
		EM       &2.05       &         -   	      &     51.8      & 0.59	&          -  		 & 58.16  		&  \textbf{0.28}&      -    & 57.55  \\
		PD       & 1.05   	 &          -	     &    52.32       &    0.46 	&           -		 & 59.75  		& \textbf{0.28}	&    -    & 57.54   \\
		DES      &  0.50 		&   7.50    &     11.47       &     1.32 	&  12.56    	& 23.52  		&  4.12 & 28.42 & 63.77  \\
		EM Daily & 0.15 	  &    13.76   &     22.62      &     0.19	&     22.23     & 41.60  	& 	0.36	&33.98 & 70.23  \\
		PD Daily   & \textbf{0.13}   &   13.78    &     22.63       &    \textbf{0.17}	& 22.23     	& 41.56   		& 0.33	&34.00& 70.18  \\
		\textbf{GP}     & \textbf{0.13}   &    \textbf{5.87}  		&    \textbf{8.13}      &   0.26  &      \textbf{8.70} 		& \textbf{14.29}	 & 	0.71& \textbf{16.38} &\textbf{31.43}  \\ \hline
		Concave   & E1 & E2  				&  E3          			& E1          	  & E2    			& E3       			& E1            & E2     & E3      \\ \hline
		EM       & 6.36       &         -   	      &     88.99     & 6.30	&   -  		 & 88.09 		&  5.90	&     -    & 83.04  \\
		PD       & 4.67  	 &          -	     &    65.33     &    4.67	&        -        & 65.34 		& 4.34	&    -    &61.12 \\		
		DES      &  0.09		&   0.78    &     1.35  &     0.22		&   1.63   	& 3.17		&  0.83 & 5.61 & 12.06 \\
		EM Daily & 0.08 	  &     0.97   &     1.31     &     0.18 	&    2.23    &3.27 		& 0.50	& 6.54 & 9.86 \\
		PD Daily   & \textbf{0.05}         &   0.76       &    1.02              &     \textbf{0.13}    	&   1.9			&  2.75  	& \textbf{0.30}	& 5.61 & 8.44  \\
		\textbf{GP}     & \textbf{0.05}    &    \textbf{0.59} 		&    \textbf{0.91}       &  0.14  &     \textbf{1.19}	& \textbf{2.13} & 	0.31 & \textbf{2.68} & \textbf{5.06} \\ \hline		
	\end{tabular}
	\caption{Results of Experiment 2. E1 denotes percentage mean error, E2 denotes average absolute daily error in the unconstrained approximation, and E3 denotes average absolute error in the total cumulative unconstrained approximation.}
	\label{table: exp 2 results}
\end{table} 

\begin{figure}
	\centering
	\begin{subfigure}{0.5\textwidth}
		\centering
		\includegraphics[width = 0.95\linewidth]{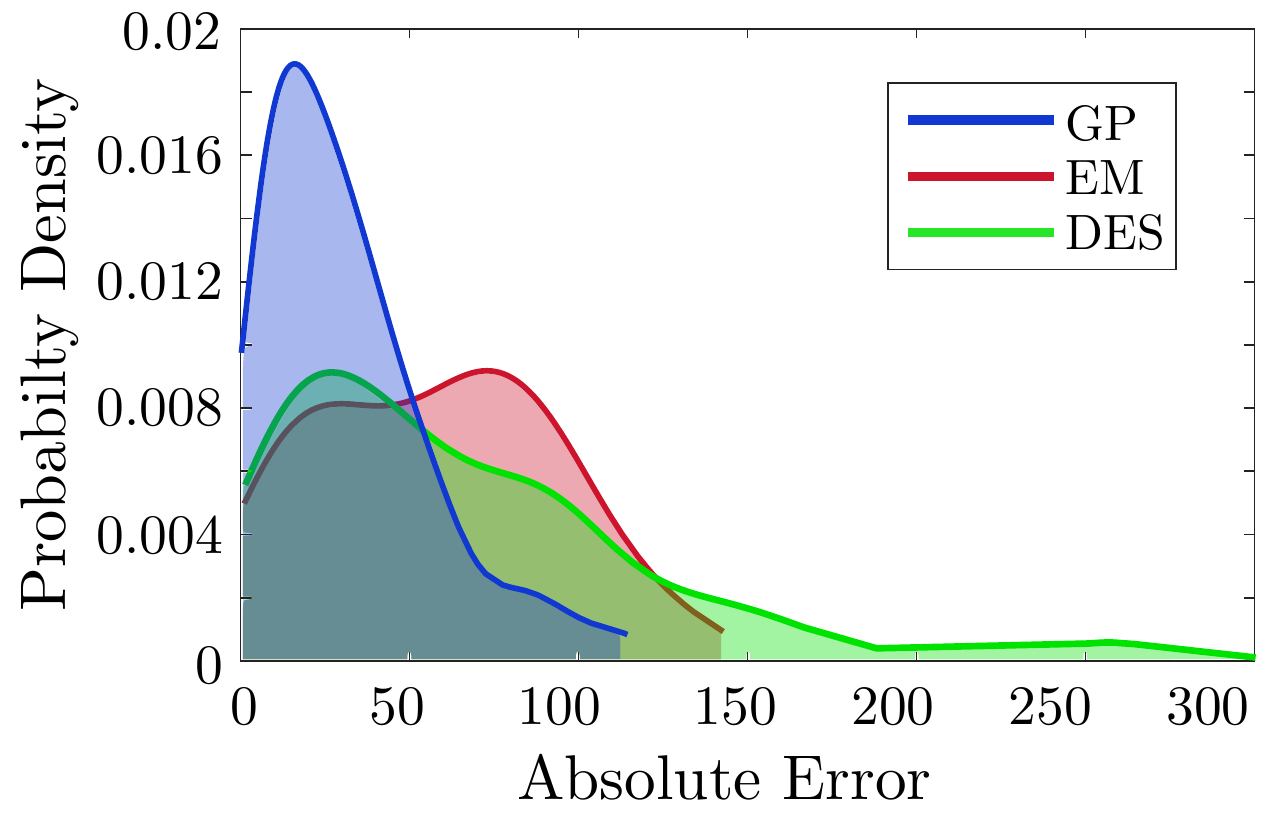}
		\caption{E3 Error probability density estimates}
		\label{Fig: exp 2 E3 error distrbutions}
	\end{subfigure}%
	\begin{subfigure}{0.5\textwidth}
		\centering
		\includegraphics[width=0.95\linewidth]{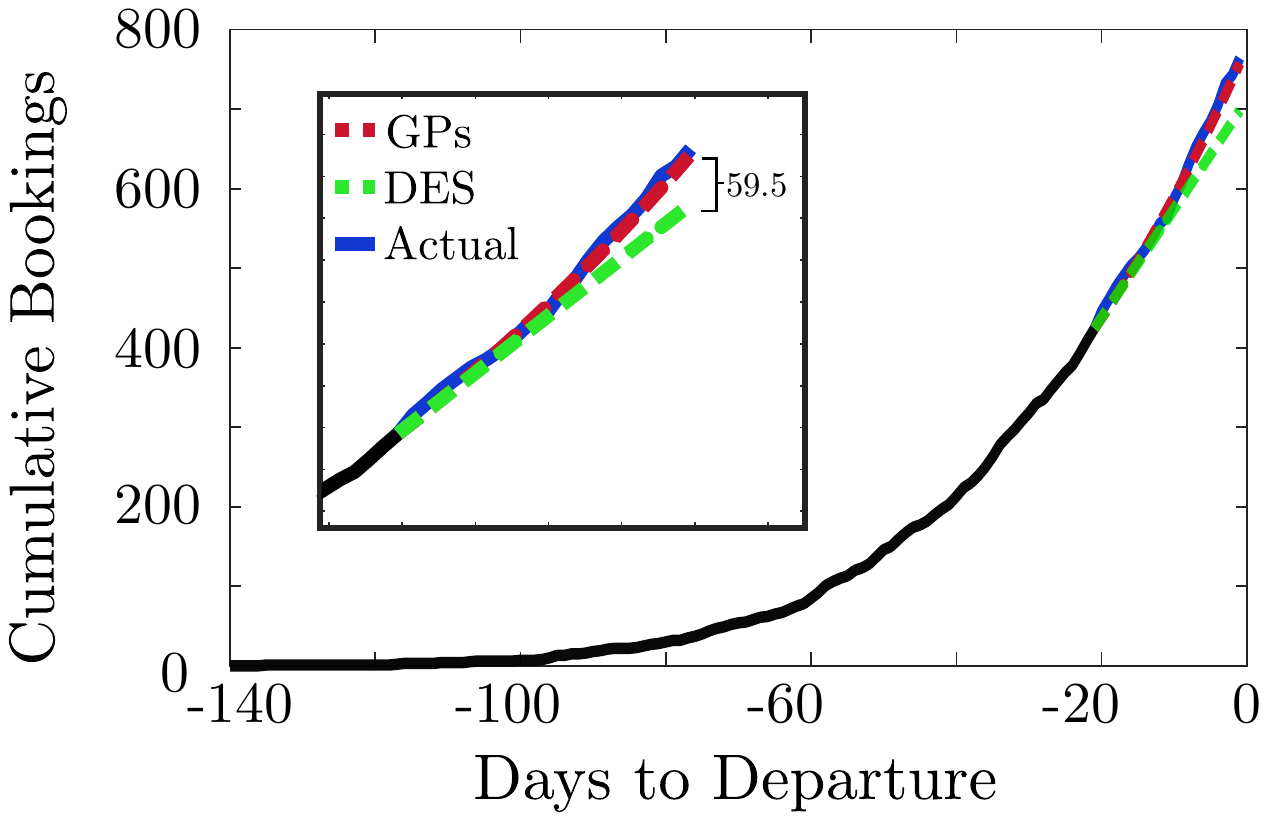}
		\caption{Extrapolation using GP and DES}
		\label{Fig: DESvsGPs}
	\end{subfigure}%
	
	\caption{Figure (a) shows the probability density estimates of E3 error totals for the convex curve set in Experiment 2, with 20 days constrained. We include only the top performing methods, and exclude the results for PD as they coincide almost precisely with those of EM. Figure (b) shows a comparison of the unconstrained extrapolations produced by GP and DES on a convex curve from Experiment 2. The black line corresponds to cumulative bookings prior to constraining.}
	
\end{figure}

From \autoref{table: exp 2 results} we see that GP regression outperforms every other method according to every measure, with four exceptions. The exceptions involve the E1 error measure on both curve sets: when the constraining period is 10 days long, E1 for GP is slightly (between 0.01\% and 0.1\%) larger than for the daily variants of EM and PD; and the same applies when 20 days were constrained, with the added outcome that E1 is also slightly lower for standard EM and PD than for GP on the convex curves. 
However, in both cases, E2 and E3 for GP is less than half what it is for all variants of EM and PD. We illustrate this stark contrast in performance in \autoref{Fig: exp 2 E3 error distrbutions}, which shows the probability density estimates of E3 error for EM, PD and GP when half the convex curve set was constrained for 20 days. These results highlight how misleading it is to consider only the percentage mean error (E1). 

Moreover, the success of EM and its variants according to the E1 error measure on the convex curves is a result of the fact that EM collapses into mean imputation when the constrained cumulative demand curves are either very steep or constrained for a substantial period of time. To see why this happens, notice that if a demand curve is constrained for a significant period of time, the observed constrained demand is likely to be significantly lower than the mean of the true demand observations.  
The first step of EM calculates the initial estimates of unconstrained demand, $\widehat{d}_i^{(0)}$, as
\begin{align}\label{eq: EM edge1}
\widehat{d}_i^{(0)} &= \E[\,d_i\,|\,d_i \geq b_i,\, d_i \sim \mathcal{N}(\mu^{(0)}, \sigma^{(0)})\,],\;\, \text{for all constrained } d_i,\\
& = \dfrac{\int_{b_i}^{\infty}x \, p(d_i=x)\,\text{d}x}{p(d_i\geq b_i)}, \;\, \text{where } d_i \sim \mathcal{N}(\mu^{(0)}, \sigma^{(0)}). \label{eq: EM to mean}
\end{align}
As $b_i$ decreases below $\mu^{(0)}-2\sigma^{(0)}$, the probability in the denominator approaches 1, and the numerator approaches $\E[d_i] = \mu^{(0)}$. EM therefore predicts every constrained value as almost exactly $\mu^{(0)}$, and since $\mu^{(0)}$ is generally taken to be the mean of the true demand observations, this is equivalent to mean imputation~\cite{weatherford2002better}. 

For example, when EM is applied to the convex curve set when curves are constrained for 20 days, it produces unconstrained approximations for all 45 constrained curves which are extremely close to the mean of the true demand totals (all within 0.3 of this value). It is unsurprising that these estimates are very close to the mean of the actual cumulative demand totals $\mathcal{A}$ given the way that the curves were constructed and how the subsets of curves to be constrained were selected (half of each curve shape). However, with the set of total demand having a standard deviation of 65, predicting the correct mean value for every unconstrained approximation results in a large absolute error on average (as shown in the E3 column of \autoref{table: exp 2 results}). The same explanation can be applied to explain the success of the daily version of EM in achieving a small E1 error, while resulting in a very large E2 and E3 error.

As expected, DES performs poorly on both convex and concave curves in this experiment, as both curve sets are constrained for the same length of time. Moreover, the increased curvature of some of the curves as compared with Experiment 1 exacerbates the failure of its linear extrapolation (see \autoref{Fig: DESvsGPs}).

\subsection{Experiment 3: `Double Poisson Process' (DPP) Data}\label{Sec: Exp 3}

To design our final experiment, we analyse a data set of 392 demand curves for Emirates Airlines tickets for several given flight routes and for individual fare-classes, with total cumulative demand above 70. In this section, we focus exclusively on demand curves of the convex type, as these are the most common in our data set. 

\begin{figure} [h] 
	\centering
	\begin{subfigure}{0.33\textwidth}
		\centering
		\includegraphics[width=\textwidth]{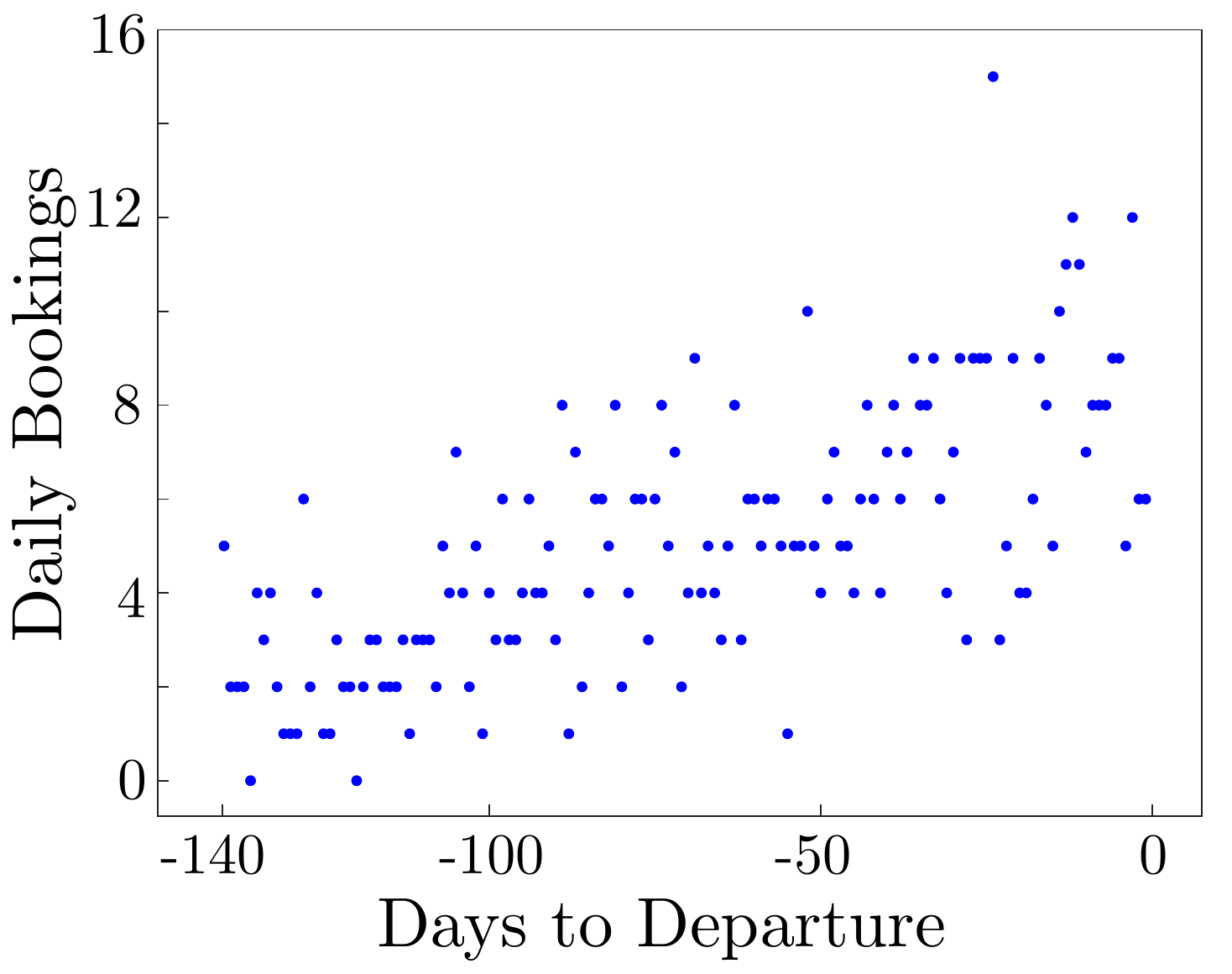}		
		\caption{Experiment 1}
		\label{Fig: Lit Data}
	\end{subfigure}%
	\begin{subfigure}{0.33\textwidth}
		\centering
		\includegraphics[width=\textwidth]{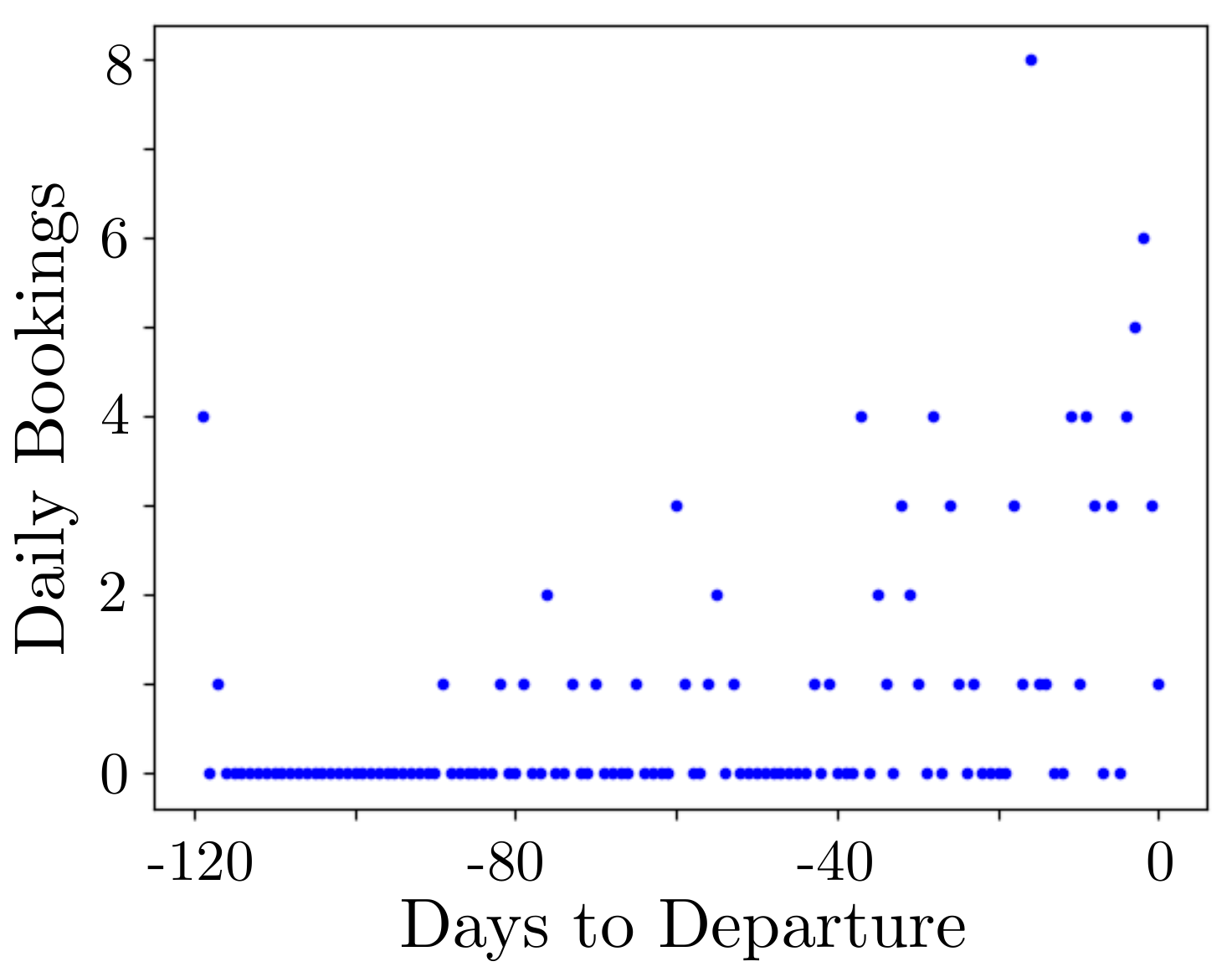}	
		\caption{Real data}
		\label{Fig: Real Data}
	\end{subfigure}%
	\begin{subfigure}{0.33\textwidth}
		\centering
		\includegraphics[width=\textwidth]{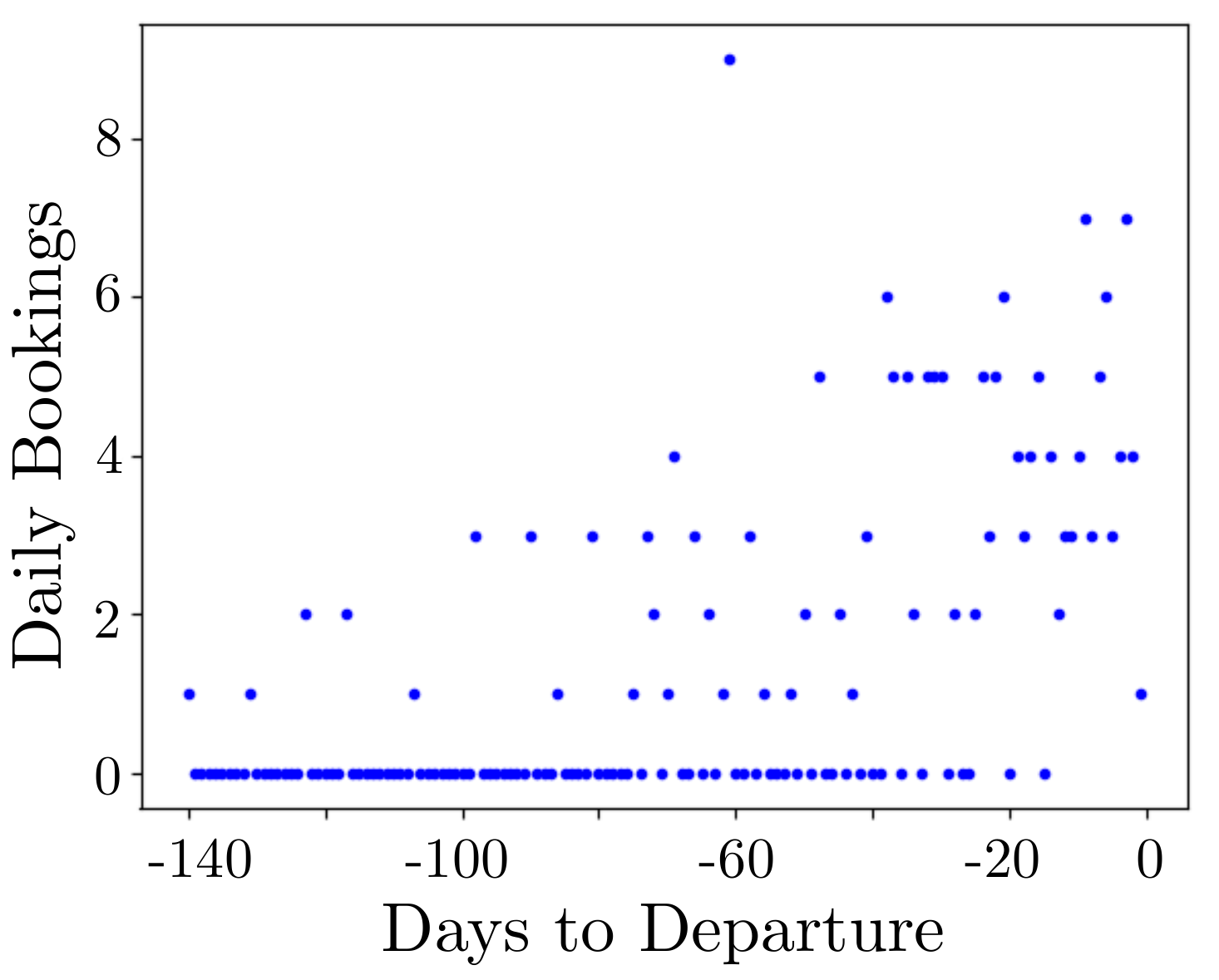}		
		\caption{Experiment 3}
		\label{Fig: DPP Data}
	\end{subfigure}
	\caption{Comparison of data generated for our experiments to real Emirates demand data. Figure~(c), generated in Experiment 3 by the `DPP' process, is notably more similar to the real data in Figure~(b) than the data created by Queenan at al.\ which is shown in Figure~(a).}
	\label{Fig: data comparisons}			
\end{figure}

\autoref{Fig: Lit Data} shows a typical example of the demand data generated for Experiment 1. The data produced for Experiment 2 is similar, except for the fact that the average daily rate increases in either a linear, quadratic, or cubic manner, depending on the demand curve in question (see the Supplementary Material for figures). While the method used in Experiments 1 and 2 to model bookings as an inhomogeneous Poisson process is standard in the literature, the resulting test data it produces is, in fact, noticeably different to the typical daily demand data we see in our data set. This is immediately apparent when comparing \autoref{Fig: Lit Data} with \autoref{Fig: Real Data}, which shows real data from a typical high demand flight. 

This motivates an alternative approach to manufacturing test data which more closely replicates typical Emirates demand data. The benefits of such a method over simply comparing the methods on the real data itself, are two-fold:
\begin{enumerate}
	\item It makes the experiment reproducible, such that it can be performed by any researcher, in particular those without access to this proprietary Emirates Airlines data set
	\item It avoids the problem of having to subset the real data into sets of demand curves which are considered to have similar demand trends (which is necessary prior to testing unordered methods, as they are only applied to such subsets). Airlines have their own ways of doing this subsetting, and the extent to which they are successful in creating appropriate subsets will greatly impact upon the performance of any unordered method. Manufacturing realistic data therefore allows us to control the bias which exists in favour of, or against, unordered methods while still producing industry-relevant results. 
\end{enumerate}

We started by attempting an alternative method for simulating an inhomogeneous Poisson process known as `thinning'. However, we found that even our best efforts to generate realistic-looking data still fell short (see the Supplementary Material for details).
As an alternative starting point for designing more realistic test data, we note that one of the obvious differences between the data in \autoref{Fig: Lit Data} and \autoref{Fig: Real Data} is that there are significantly more $0$ data points (days on which $0$ bookings were made) in the latter figure, and these continue until much closer to the day of departure. This motivates an approach which independently models two key features of the data:

\begin{enumerate}
	\item The rate $\lambda(t)$ representing the trend in average daily bookings (on those days on which bookings are made);
	\item  The `inter-arrival times' (on the scale of days), that is, the number of days with zero bookings in between days on which bookings are made. 
\end{enumerate} 
Modelling these two factors independently does indeed break the assumption of an underlying standard Poisson arrivals process. However, we find that doing so allows us to achieve more realistic-looking data nonetheless. 

We propose sampling both `arrivals' (bookings) and the `inter-arrival times' (the number of days between bookings) from Poisson distributions, with time dependent rates $\lambda_1(t)$ and $\lambda_2(t)$ respectively.\footnote{We note that the Poisson distribution is an especially unusual choice for modelling inter-arrival times, since it means that they must be integer valued. However, since we are operating on the scale of `days', and producing a data set which gives daily bookings, this property of the Poisson distribution does not cause any problems: we hope to model precisely the integer number of days between days on which bookings occur.} To create a test curve, we therefore define two vectors $\boldsymbol{\lambda_1}, \boldsymbol{\lambda_2} \in \mathbb{R}_+^{140}$, with each element corresponding to the rate at each day in the lead up to departure. Thereafter we follow two steps: we use the vector $\boldsymbol{\lambda_2}$ to determine on which days non-zero bookings occur, and on each of these days we sample the number of bookings from a Poisson distribution with mean rate equal to the element of $\boldsymbol{\lambda_1}$ corresponding to that day (see Figure 3 in the Supplementary Material for an illustration of this process).

The data shown in \autoref{Fig: DPP Data} was generated with this `double Poisson process' (DPP), and it is evidently much more similar to those in the Emirates data set. We thus use this process to generate 90 convex demand curves of varying degrees of convexity as in Experiment 2 (the formulae for $\lambda_1(t)$ and $\lambda_2(t)$ used to create these curves are given in the Supplementary Material). The resulting curves have a mean total demand of 181.9 and a standard deviation of  35.2. We repeat this process three times, constraining 45 of the curves for 5, 10 and 20 days, and compare the performance of the various unconstraining methods in each case. As before we repeat this whole process five times, and we present the averaged results in \autoref{table: DPP convex results}.

\begin{table}[]
	\centering
	\small
	\begin{tabular}{p{3.3cm}K{1cm}K{1cm}K{1cm}|K{1cm}K{1cm}K{1cm}|K{1cm}K{1cm}K{1cm}}
		\hline
		& \multicolumn{9}{c}{No. of Days Constrained}                                              \\ \hline
		& \multicolumn{3}{c|}{5 Days} & \multicolumn{3}{c|}{10 Days} & \multicolumn{3}{c}{20 Days} \\ \hline
		& E1    & E2    & E3     & E1     & E2    & E3     & E1    & E2     & E3     \\ \hline
		EM                & 7.06  & -     & 30.04  & 3.22   & -     & 26.45  & 1.10  & -      & 29.08  \\
		PD                & 4.51  & -     & 26.88  & 1.48   & -     & 26.57  & 0.86  & -      & 29.20  \\
		DES               & 0.98  & 4.02  & 6.14   & 2.64   & 6.93  & 12.30  & 8.15  & 13.88  & 30.81  \\
		EM Daily          & 0.59  & 5.32  & 8.65   & 1.15   & 7.67  & 13.69  & 2.30  & 12.27  & 24.58  \\
		PD Daily          & \textbf{0.31}  & 5.05  & 8.12   & 0.44   & 7.22  & 12.78  & \textbf{0.49}  & 10.94  & 22.36  \\
		\textbf{GP}               & \textbf{0.31}  & \textbf{3.39}  & \textbf{4.88}   & \textbf{0.40}   & \textbf{5.60}  & \textbf{9.19}   & 1.46  & \textbf{9.56}   & \textbf{18.65} \\ \hline
	\end{tabular}
	\caption{Results of Experiment 3. As in Table \ref{table: exp 2 results}, E1 denotes percentage mean error, E2 denotes average absolute daily error in the unconstrained approximation, and E3 denotes average absolute error in the total cumulative unconstrained approximation.}
	\label{table: DPP convex results}
\end{table}

Once again, GP regression outperforms all other methods, in every case and by almost every measure. The only exception is that we see a repeat of the result we saw in Experiment 2, where the E1 error for a constrained period of 20 days is slightly higher for GP than it is for EM, PD and PD Daily. However, GP is again noticeably better according to both the E2 and E3 error measures, which we argue are also relevant for evaluating the performance of the various methods, for reasons discussed in \autoref{Section: Exp 1}.

\subsection{Experimental Conclusions}
The results from all three experiments described in this section indicate that our proposed GP unconstraining method achieves significantly better results on average than existing state of the art methods in the literature.

\section{Detecting Changepoints in Demand Trends}\label{Chapter: Changepoints}

Thus far we have assumed that the underlying trend in demand is likely to be generally smooth, without kinks and discontinuities, and we have compared the performance of GPs to existing unconstraining methods using test data which stays true to this assumption. In this section we weaken this smoothness assumption, allowing for points at which the characteristics of the underlying demand trend change dramatically. 

In time-series analysis, times at which the characteristics of the data change dramatically are known as changepoints. Research on methods for detecting changepoints has been ongoing for decades \cite{hinkley1970inference, basseville1993detection}, and a range of methods have been applied to data from a diverse set of subjects ranging from hydrology \cite{wong2006change} to the history of political relationships \cite{western2004bayesian}. 

Demand for flight tickets is affected by many exogenous factors, some of which (like a competitor's prices or the perceived safety of a destination) can change drastically in short spaces of time. This makes it likely that changepoints will feature frequently in the demand data, and we have observed them in the Emirates data discussed in \autoref{Sec: Exp 3}. Any method which hopes to accurately pick up and forecast a demand trend based on past time-series data will make significant errors unless it is able to account for the possibility of changepoints. Without registering a changepoint, the method's extrapolation will be informed by all past data, when in fact it needs to account for the fact that the data from before the changepoint does not accurately represent the demand trend after it. 

Garnett at al.\ \cite{garnett2009sequential} showed how changepoint detection can be simply and elegantly incorporated into a GP regression framework by constructing an appropriate covariance function. For our purposes, we want to allow for the fact that the covariance before and after the changepoint might be completely different (for example, a roughly quadratically increasing demand level before the changepoint, and a small, approximately constant demand level after the changepoint).
We therefore define our covariance function to be
\begin{align}\label{eq: Changepoint covariance}
k(x,x') = \begin{cases}
\sigma_1^2(x^\top x' + c_1)^{p_1} & \text{if } x,x' < x_c,\\
\sigma_2^2(x^\top x' + c_2)^{p_2} & \text{if } x,x' \geq x_c,\\
0 & \text{otherwise},
\end{cases}
\end{align}
where $x_c$ is the location of the change point, and is also a hyperparameter which is inferred from the data along with the other covariance function hyperparameters $\theta_c = \{\sigma_1, \sigma_2, c_1, c_2, p_1, p_2, x_c \}$. We implement this by adding a further covariance function file to the GPML library, and we marginalise out all hyperparameters using the quadrature method presented at the end of \autoref{Section: GP regression}. 

\begin{figure*}[!p]
		\centering
		\begin{subfigure}{0.5\textwidth}
			\centering
			\includegraphics[width=0.9\textwidth]{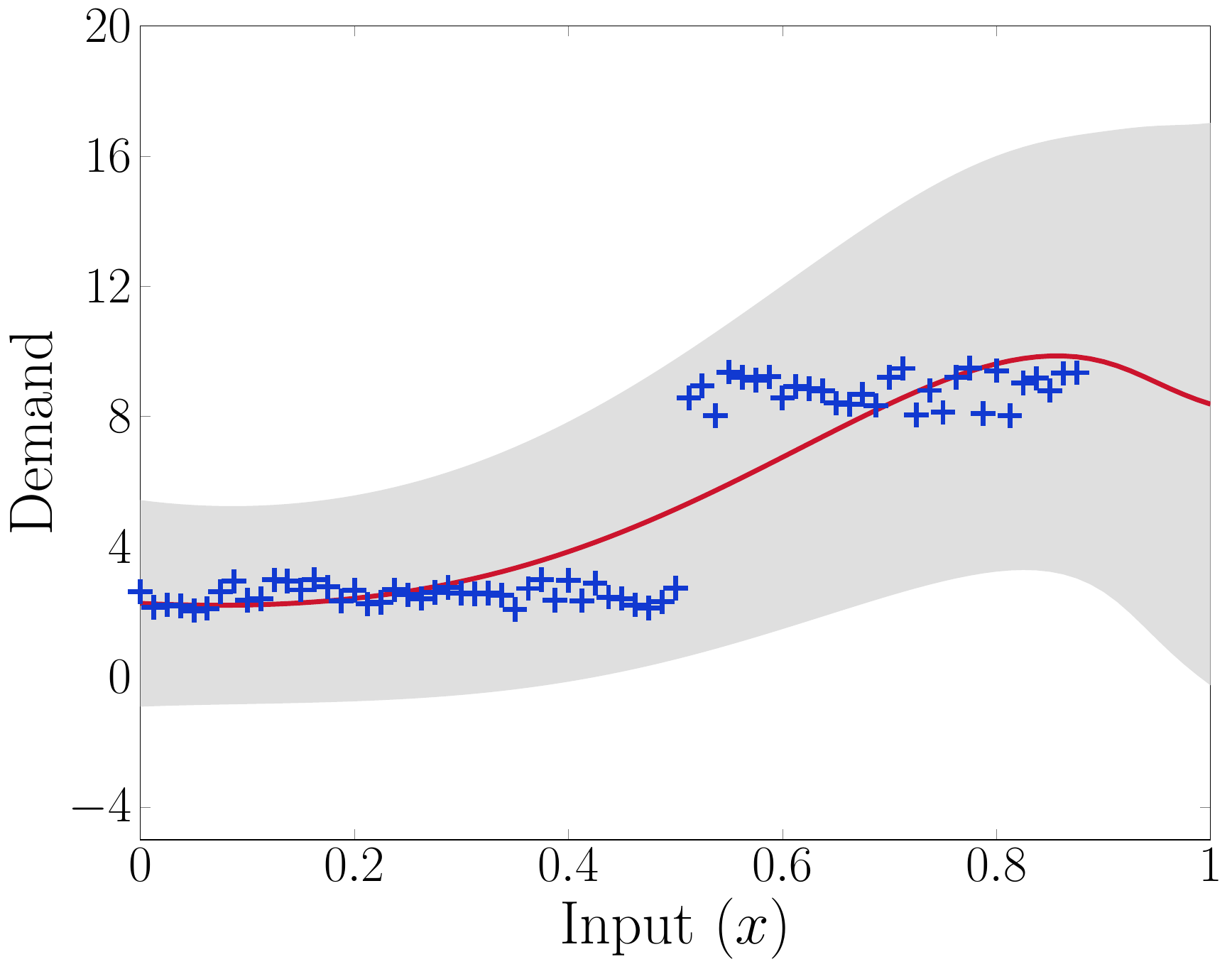}		
			\caption{Scenario 1, standard covariance}
			\label{Fig: frf std}
		\end{subfigure}%
		\begin{subfigure}{0.5\textwidth}
			\centering
			\includegraphics[width=0.9\textwidth]{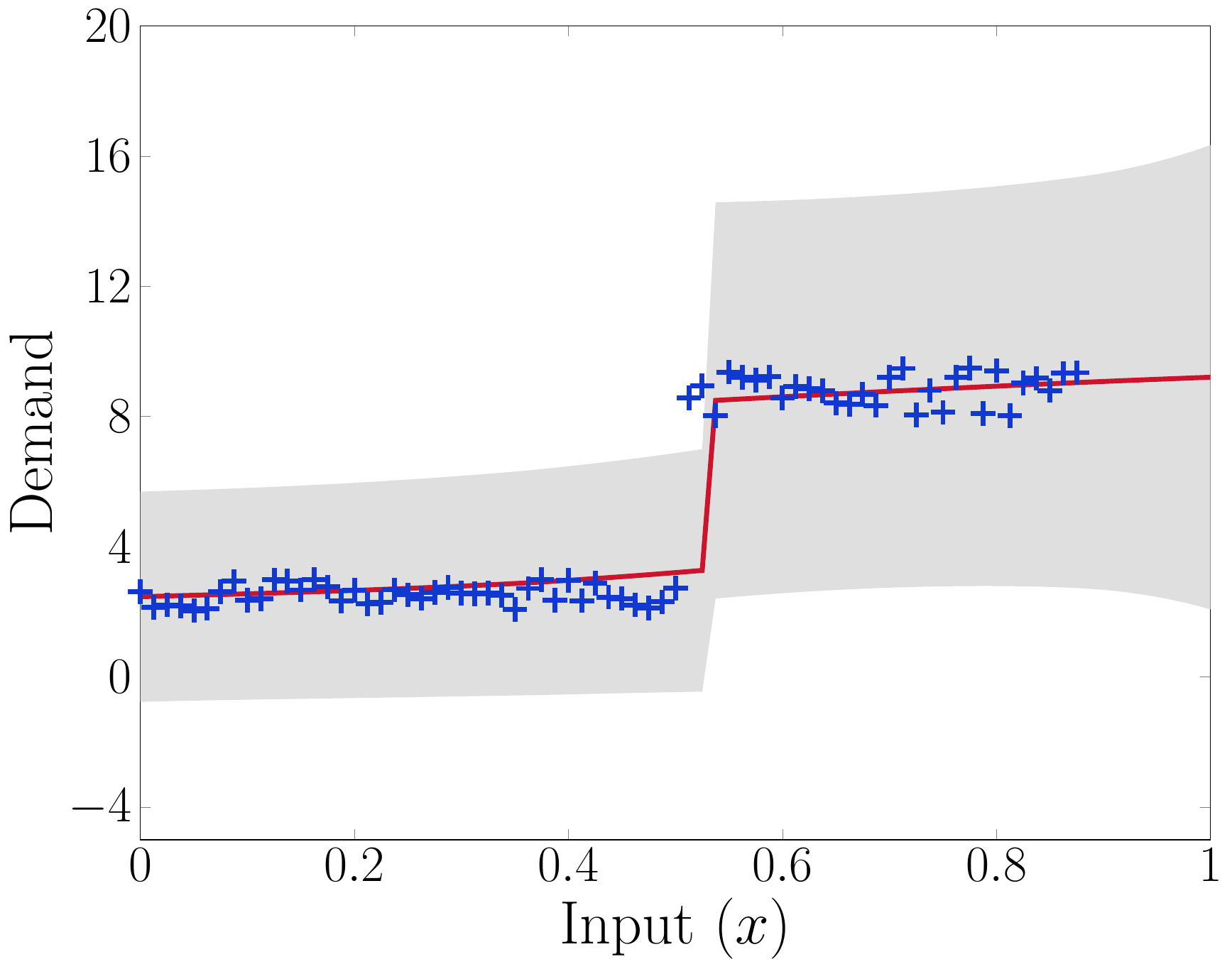}		
			\caption{Scenario 1, changepoint covariance}
			\label{Fig: frf cp}
		\end{subfigure}
		\begin{subfigure}{0.5\textwidth}
			\centering
			\includegraphics[width=0.9\textwidth]{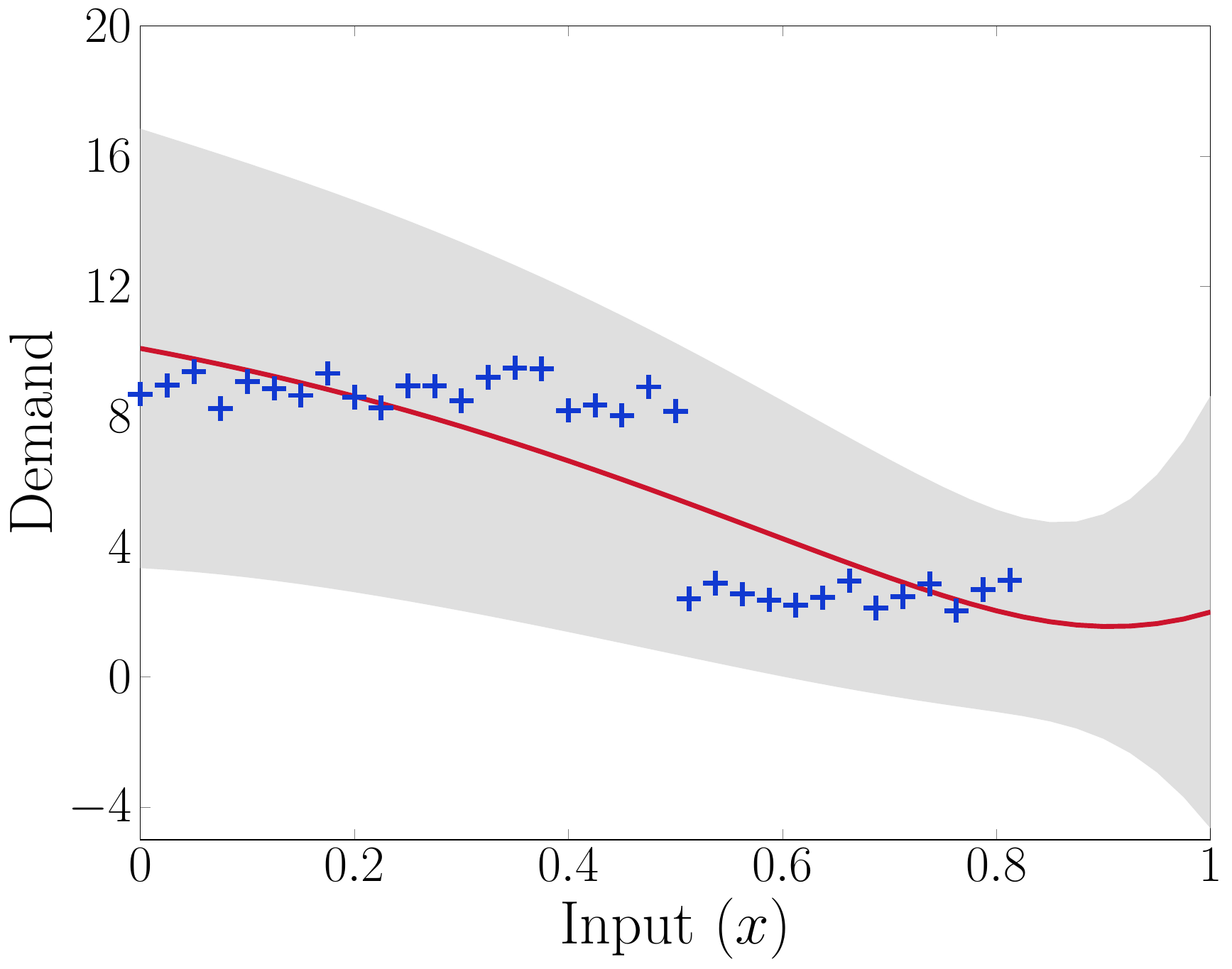}		
			\caption{Scenario 2, standard covariance}
			\label{Fig: fdf std}
		\end{subfigure}%
		\begin{subfigure}{0.5\textwidth}
			\centering
			\includegraphics[width=0.9\textwidth]{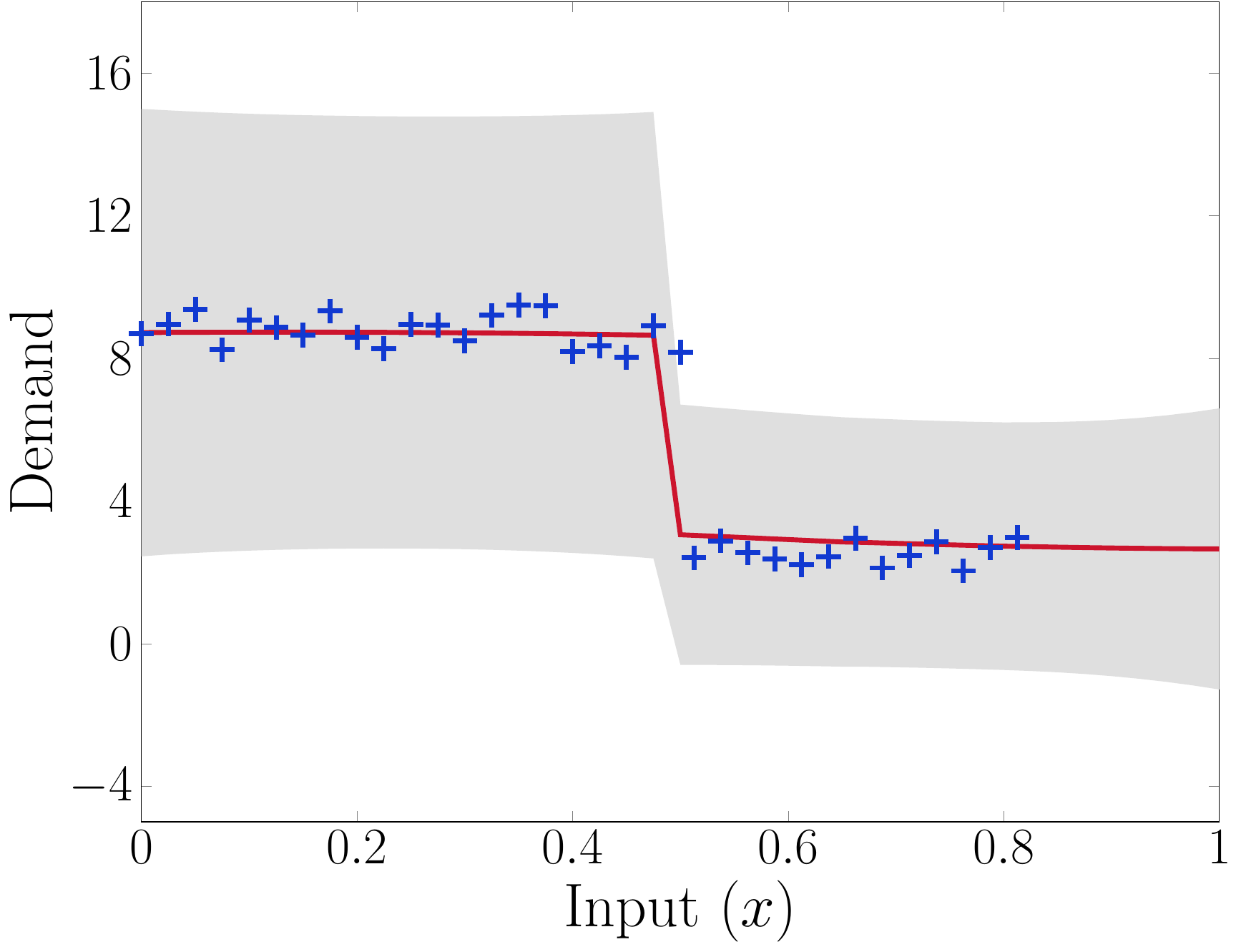}		
			\caption{Scenario 1, changepoint covariance}
			\label{Fig: fdf cp}
		\end{subfigure}
		\begin{subfigure}{0.5\textwidth}
			\centering
			\includegraphics[width=0.9\textwidth]{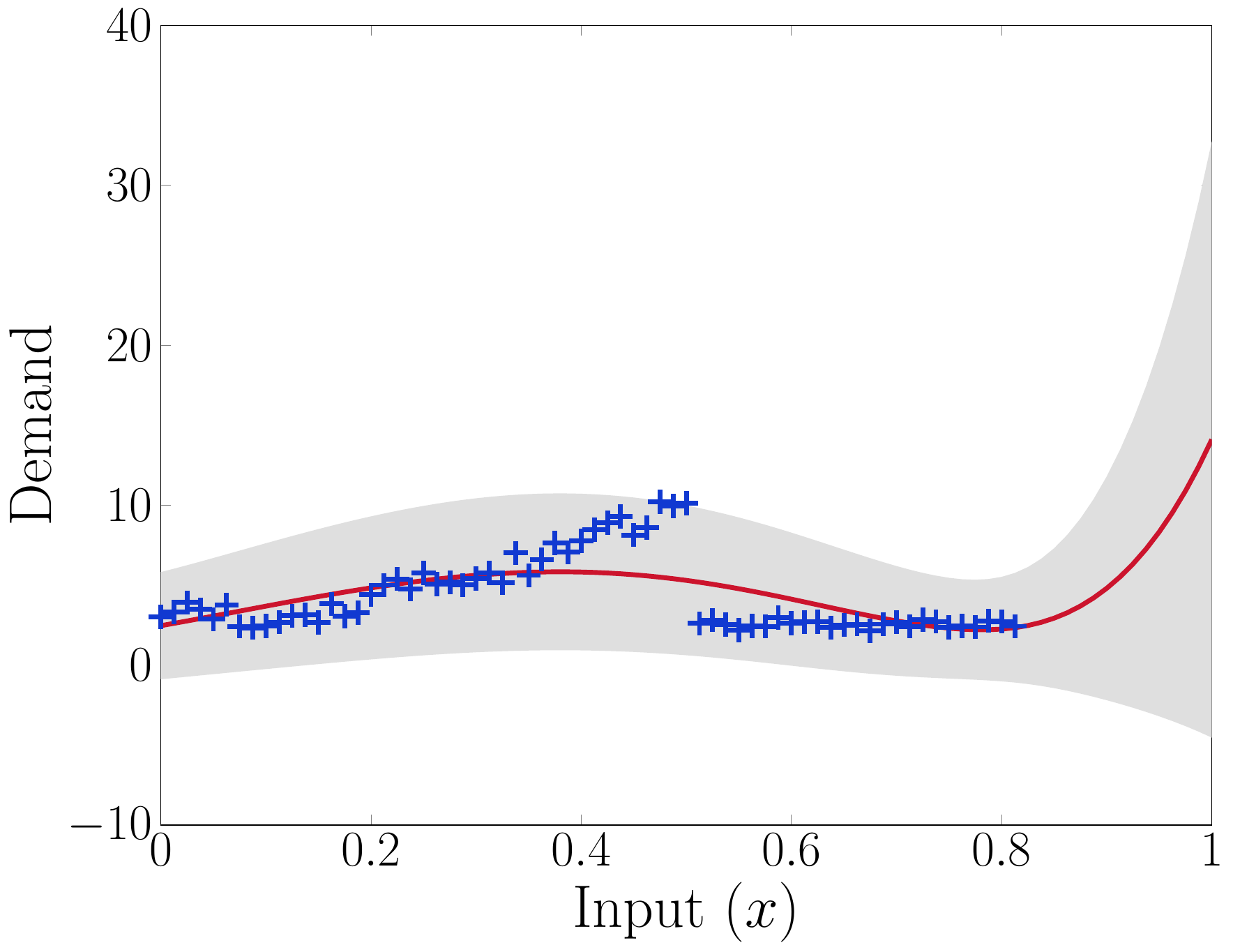}		
			\caption{Scenario 3, standard covariance}
			\label{Fig: pdf std}
		\end{subfigure}%
		\begin{subfigure}{0.5\textwidth}
			\centering
			\includegraphics[width=0.9\textwidth]{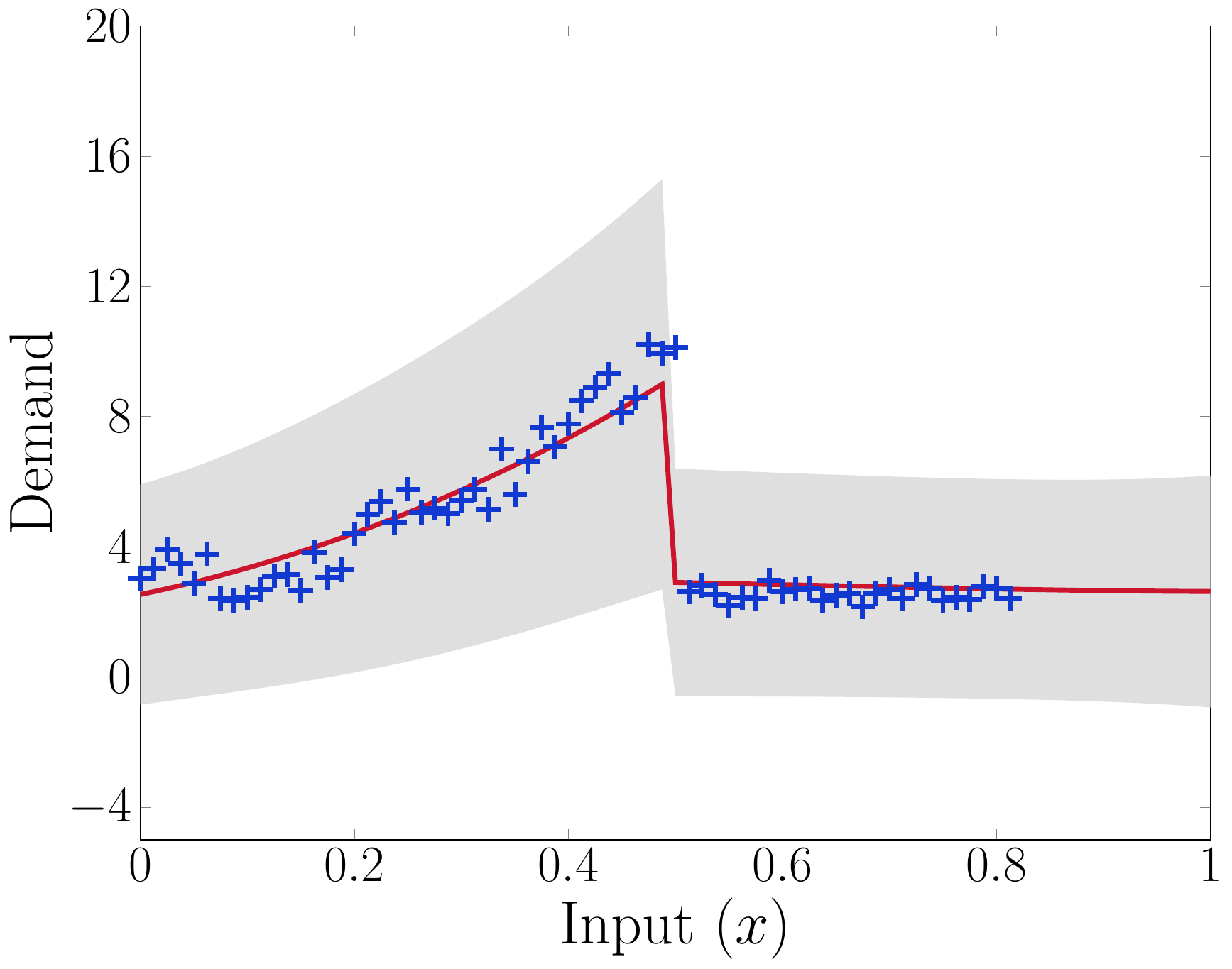}		
			\caption{Scenario 3, changepoint covariance}
			\label{Fig: pdf cp}
		\end{subfigure}
		
		\caption{Prediction and extrapolation in the presence of changepoints. In each scenario, the data is the same in the figures on the left and right. The figures on the left show the GP fit and extrapolation using the covariance function given in \eqref{eq: our cov func}, while the figures on the right use the modified changepoint covariance function defined in \eqref{eq: Changepoint covariance}.}
		\label{Fig: CP scenarios}			
	\end{figure*}

We omit comparisons with other methods in our analysis of changepoint detection. Since detecting changepoints is exclusively applicable when demand data is viewed as time-series data, direct comparisons with unordered methods like EM and PD do not make sense. Moreover, the fundamental problem with DES (that it is limited to linear extrapolation of cumulative demand) would apply equally in the presence of changepoints. Instead, we illustrate the power of our method with a number of scenario-inspired test cases, with normalised inputs ($x\in [0,1]$), the results of which are shown in \autoref{Fig: CP scenarios}. 
In particular, we consider the demand for a given fare-class on a given flight on Airline A in the following three scenarios:
\begin{itemize}
	\item Scenario 1 (Figures \ref{Fig: frf std} and \ref{Fig: frf cp}): Demand is fairly constant but relatively low because the price of the ticket is not competitive. When a competitors' price suddenly increases or their flights are sold out, this diverts demand to Airline~A, causing a jump in the average daily bookings. 
\end{itemize}

	\begin{itemize}
	\item Scenario 2 (Figures \ref{Fig: fdf std} and \ref{Fig: fdf cp}): The inverse of Scenario 1, where Airline A's flight begins with high relatively constant demand, until a cheaper option on a competitor airline becomes available, diverting demand away from Airline A. 
	\item Scenario 3 (Figures \ref{Fig: pdf std} and \ref{Fig: pdf cp}): Demand for Airline A's flight is increasing in some non-linear fashion until there is some destination-related shock (a headline news item causing the destination to be considered unsafe, for example), after which demand drops dramatically, and remains fairly flat at a relatively low level.
\end{itemize}

The plots on the left-hand side (Figures \ref{Fig: frf std}, \ref{Fig: fdf std}, and \ref{Fig: pdf std})  show the results of fitting our proposed GP regression model with the standard covariance function given in \eqref{eq: our cov func}, which does not account for changepoints. We compare these with the plots shown on the right-hand side (Figures \ref{Fig: frf cp}, \ref{Fig: fdf cp}, and \ref{Fig: pdf cp}) which are obtained using the changepoint covariance function defined in \eqref{eq: Changepoint covariance} on the same data. 
It is clear in all cases that the changepoint covariance function better extrapolates the post-changepoint demand trend, and Figure \ref{Fig: pdf std} best exhibits just how wrong the GP prediction might become when a changepoint is not accounted for. In this case, demand has clearly collapsed to a low level, and there is no indication that it is likely to pick up again. The changepoint covariance function picks this up and correctly extrapolates this low, relatively constant demand. When the changepoint is not accounted for however, the predicted future demand begins to increase rapidly in strong contrast to the post-changepoint trend. 

For illustration purposes, we have  thus far only included examples of data with a single changepoint.\footnote{Though the existence of a changepoint is pre-specified when defining the changepoint covariance function \eqref{eq: Changepoint covariance}, if it transpires that there is in fact no changepoint, the inferred covariance parameters on either side of $x_c$ will be roughly the same, resulting in a prediction similar to that made by the standard, non-changepoint covariance function given in \eqref{eq: our cov func}.} However, our approach can easily be extended to cope with multiple changepoints given sufficient computational power (Garnett at al.\ \cite{garnett2009sequential} show that in fact, when taking a `moving window' approach to changepoint detection, it is uncommon to need to account for more than two changepoints in the covariance function itself). This ability to detect and account for discontinuities in the demand trend is a powerful motivating factor for the use of our method for unconstraining demand.

\section{Conclusions and Future Work}
In this paper we proposed and extensively tested a new single-class unconstraining method that uses GP regression with a Poisson likelihood and a new covariance function --- a `variable degree polynomial covariance function'. This proposed regression model is novel in its inference of the (non-integer) exponent in the polynomial covariance function, and its ability to perform well on more realistic demand data; data which exhibits nonlinear inhomogeneous Poisson rates, varying inter-arrival-times and discontinuities in demand. 

The results of our numerical experiments point to a number of important conclusions. The first is that it is important to consider not only the percentage mean error (as is standard in the literature), but also the average absolute error in the unconstrained approximations. For some applications of unconstraining --- say, in estimating average historical demand for pricing and route-planning purposes --- accurately estimating the mean of historical demand may well be the most important goal. However, the decision about whether or not to re-open a currently-constrained fare-class depends on accurately unconstraining that particular demand curve, not knowledge of the historical mean, and hence what matters in such cases is the accuracy of individual unconstrained approximations. Importantly, our results show that success according to one measure does not imply success according to the other, and hence in general we ought to consider both.
Secondly, our results show that when both error measures are considered, our proposed GP method notably outperforms all other unconstraining methods included in our experiments. Our GP method performs comparably (or better) when considering the percentage mean error, and performs decidedly the best when considering average absolute error. One notable drawback was that with un-vectorised code, the use of GPs is significantly more computationally expensive than other methods. However, we expect that with vectorised and optimised code, GPs would be much more computationally comparable.

Our final piece of analysis extends our proposed method to cope with changepoints in time-series demand data, by defining an amended covariance function, and inferring both the location of the changepoint and the covariance hyperparameters on either side of it from the data. We show that extrapolation based on data which exhibits a changepoint is much more accurate when the amended (changepoint) covariance function is used. 

While these results are very positive, they also highlight areas which merit further research. First, it would be preferable (though non-trivial) to find a set of conditions under which we can guarantee that the variable degree polynomial covariance function we propose will work with a predetermined spectral shift. Secondly, though we find GPs to be preferable to all other methods tested in our experiments, the superior performance of EM Daily and PD Daily in Experiments 1 and 3, as compared with the standard average-based EM and PD, merits further investigation by RM practitioners into increasing the granularity of the data on which they apply these methods.

\section*{Acknowledgements}
\noindent This work was supported by the Oxford-Emirates Data Science Lab.

\section*{References}

\bibliographystyle{model5-names}
\bibliography{mybibfile.bib}

\end{document}